\documentclass[12pt]{article}
\pdfoutput=1
\usepackage{amsmath,amssymb,amsfonts,amscd,bm,marginnote,cite,graphicx}
\usepackage[colorlinks=true, pdfstartview=FitV, linkcolor=black, citecolor=black, urlcolor=black]{hyperref}
\usepackage[usenames,dvipsnames]{xcolor}
\newcommand{\hlt}[1]{{\color{WildStrawberry}{\em #1}}\index{#1}}

\edef\marginnotetextwidth{\the\textwidth}

\newcommand{\thistitle}{
Quantum Spaces are Modular}
\newcommand{\addresspi}{
	Perimeter Institute for Theoretical Physics, 
	31 Caroline St. N.,  Waterloo ON, N2L 2Y5, Canada
	}
\newcommand{\addressuiuc}{
	Department of Physics, University of Illinois,
 	1110 West Green St., Urbana IL 61801, U.S.A.
	}
\newcommand{\addressvt}{
	Department of Physics, Virginia Tech,  
	Blacksburg VA 24061, U.S.A.
	}
\newcommand{\emaillf}{lfreidel@perimeterinstitute.ca}
\newcommand{\emailrgl}{rgleigh@illinois.edu}
\newcommand{\emaildm}{dminic@vt.edu}

\newcommand{\bra}[1]{\langle #1 |}
\newcommand{\ket}[1]{| #1 \rangle}

\newcommand{\Heis}{H}

\newcommand{\be}{\begin{equation}}
\newcommand{\ee}{\end{equation}}
\newcommand{\beq}{\begin{eqnarray}}
\newcommand{\eeq}{\end{eqnarray}}
\newcommand{\bea}{\begin{eqnarray}}
\newcommand{\eea}{\end{eqnarray}}
\newcommand{\beqn}{\begin{eqnarray}}
\newcommand{\eeqn}{\end{eqnarray}}

\newcommand{\X}{\mathbb{X}}
\newcommand{\Y}{\mathbb{Y}}
\newcommand{\K}{\mathbb{K}}
\newcommand{\Pm}{\mathbb{P}}

\newcommand{\comment}[1]{}

\def\pa{\partial}

\newcommand{\rd}{\mathrm{d}}
\def\dd{\!\cdot \!}

\def\bz{{\bar{z}}}
\def\dd{\!\cdot \!}

\def\Ph{\cal{P}}

\def\bra{\langle}
\def\ket{\rangle}

\def\vp{\varphi}

\def\dd{\!\cdot \!}

\def\Ph{{\cal{P}}}

\def\bra{\langle}
\def\ket{\rangle}

\def\PP{\mathbb{P}}
\def\tx{\tilde{x}}
\def\ty{\tilde{y}}
\def\tn{\tilde{n}}
\newcommand{\myfig}[3]{
	\begin{figure}[ht]
	\centering
	\includegraphics[width=#2cm]{#1}\caption{#3}\label{fig:#1}
	\end{figure}
	}

\renewcommand{\thefootnote}{\fnsymbol{footnote}}
\begin{document}

\title{\thistitle}


\author{
	{Laurent Freidel,$^{a}\footnote{\emaillf}$\; Robert G. Leigh$^{b}\footnote{\emailrgl}$\; and Djordje Minic$^{c}$\footnote{\emaildm}}\\
	\\
	{\small ${}^a$\emph{\addresspi}}\\ 
	{\small ${}^b$\emph{\addressuiuc}}\\ 
	{\small ${}^c$\emph{\addressvt}}\\
\\}
\maketitle\thispagestyle{empty}
\vspace{-5ex}
\begin{abstract}
At present, our notion of space is a classical concept. 
Taking the point of view that quantum theory is more fundamental than classical physics, and that space should be given a purely quantum definition, we revisit the notion of Euclidean space from the point of view of quantum mechanics.
Since space appears in physics in the form of labels on relativistic fields or  
Schr\"odinger wave functionals, we  propose to define  Euclidean  quantum space as a choice of polarization for the Heisenberg algebra of quantum theory.
We show, following Mackey, that generically, such polarizations contain a fundamental length scale
and that contrary to what is implied by the Schr\"odinger polarization, they  possess topologically distinct spectra. These are the modular spaces. We show that they naturally come equipped with additional geometrical structures usually encountered in the context of string theory or generalized geometry. Moreover, we show how modular space reconciles the presence of a fundamental scale with translation and rotation invariance. We also discuss how the usual classical notion of space comes out as a form of thermodynamical limit of modular space while the Schr\"odinger space is a singular limit.
\end{abstract}
\bigskip

\setcounter{footnote}{0}
\renewcommand{\thefootnote}{\arabic{footnote}}

The concept of classical space-time is one of the basic building blocks of physics.
It has evolved dramatically from Euclid to Newton and then from Minkowski to Einstein.
Moreover, since the advent of Einstein's theories of relativity, 
the concept of locality in a classical space-time has become one of the cornerstones of modern physics.
It is one of the key properties underlying effective field theory, which is widely considered a fundamental tool 
in describing various physical phenomena, capturing the main features of disparate physical systems at low energy scales.
What is very surprising in these developments is the fact that 
quantum physics has exerted little or no influence on our views about space and time.
If one excludes quantum gravity, all theories assume that quantum processes occur in the space-time of classical physics.
Yet we know that non-locality is one of the most counter-intuitive and central characteristics of quantum mechanics, and thus
our current picture of space-time in fundamental quantum theories is likely incomplete.

That non-locality must enter any theory of quantum gravity is guaranteed by the presence of a fundamental length scale, the Planck length.  We also expect that in any theory of quantum gravity the classical concept of space-time must be replaced by some suitable quantum notion. Therefore the open challenge underlying any theory of quantum gravity is  to understand the nature of quantum space-time. The usual view, however, is that  questions of quantum space-time should be tied  to gravitational phenomena involving strong curvature in some way, and/or that the modifications of quantum space-time are restricted to regions of space-time of the size of the Planck length. These would seem to be natural conclusions from the point of view of local effective field theories. One expects that aspects of quantum space-time are irrelevant to ordinary non-gravitational phenomena.

In this paper, we challenge this perspective and argue that one should revisit our current concepts of space and time in the light of quantum mechanics, even before establishing a full theory of quantum gravity. We argue that quantum mechanics itself allows for the introduction of such new concepts. The evolution of our thinking on this subject has grown from the realization that metastring theory \cite{Freidel:2013zga, Freidel:2014qna, Freidel:2015pka, Freidel:2015uug} gives rise to a quantum structure that we call {\it modular space-time}. In the present paper, we focus on space only by retreating to ordinary quantum mechanics and show that a notion of modular space is already present in the theory, and has in fact had a number of familiar applications. We regard this discussion as laying the groundwork for a more complete understanding of modular space-time.  

Ultimately, what we are interested in are {\it quantum systems with a built-in length scale}. What we will argue here is that our ability to understand such systems has been diminished by precisely the assumption that quantum physics takes place in a classical space. Philosophically though, one might expect that classical notions such as space are determined not {\it a priori}, but by the nature of the {\it probes} of a quantum system. Indeed there is ample evidence that the apparent physics, for example of the phase of a material, is determined by probe properties such as energy \cite{Davis2007,Davis2010}. And in gravitational physics, there is recent strong evidence that the observer plays a fundamental role\cite{Strominger:2014pwa, Donnelly:2016auv}. 

One of the key ideas we present here is that the usual classical notion of space is not a necessity. If we accept the idea that probes are in general quantum, we will see that a more general notion of space can emerge. 
The idea is to simply view space as a choice of polarization in phase space. We will show that there are quantum polarizations that do not have classical analogs. These polarizations are generic at the quantum level and the usual Schr\"odinger representation of the Heisenberg group that corresponds to classical space is obtained only in a degenerate limit in which notions of a fundamental  length scale are washed away.

The primary conceptual problem in understanding the physics of modular space-time is in the nature of time itself. 
This opens up new and fascinating questions about what could be the meaning of generalizing the usual Schr\"odinger evolution to modular time. We will consider such questions elsewhere. 

The present paper is organized as follows. 
First, we carefully discuss the Heisenberg group and its representations. We find that generally quantization corresponds to identifying a commutative subgroup. The usual examples correspond to classical Lagrangian submanifolds of phase space, but generically, representations correspond to {\it quantum (or modular) polarizations} which we recognize as modular spaces. These modular spaces are compact cells in phase space and carry units of symplectic flux. The construction should be recognized as a formalization of Aharonov's discussion of modular  variables in quantum mechanics \cite{Aharonov}. Next we introduce the Zak transform which maps unitarily between the Schr\"{o}dinger representation and the modular polarizations. The modular polarizations are determined by a choice of lattice within phase space as well as a co-cycle, which we relate to the existence of a symmetric bilinear form $\eta$ of split signature. Equivalently, the co-cycle can be thought of as related to the existence of a connection with integral flux through an elementary lattice cell. In the context of modular polarizations, we show that there is generally no well-defined notion of translation invariant vacuum. Instead, we show that a vacuum can be identified with a state that is annihilated by a kinetic ``Hamiltonian". This ``Hamiltonian" is associated with a second symmetric bilinear form $H$ on phase space. Thus we are led to the conclusion that general quantizations correspond to a choice of three structures, $(\omega, \eta, H)$ where $\omega$ is the symplectic structure that defines the Heisenberg algebra. Parenthetically, these structures are precisely the geometrical structures found in the simplest metastring theory, and hence we believe that metastring theory can be identified with a quantum theory of space-time, which is apparently fully consistent. In a final section, we explain the resolution of the symmetry paradox: since a modular quantization is associated with a discrete lattice, naively it appears that rotational and translational symmetry are broken. The paradox is resolved by recognizing that an observer's notion of space corresponds to a choice of basis (a choice of polarization), and such a notion is acted upon by rotations. However, precisely because the underlying algebra is non-commutative, any such choice is unitarily equivalent to any other, and thus the action of rotations induces a {\it superposition of modular spaces}. 
This reasoning leads us to the notion of {\it extensification}, corresponding to the emergence of a classical space. An extensification can be thought of as a many-body effect (or more precisely, a many-observer effect). Technically, it arises by taking the flux, alluded to above, to infinity in a coherent way.

\section{Modular Operators and Quantum Mechanics}

\subsection{Quantum space}

The notion of space and the notion of time pervades every other concept in physics. They are of course central concepts in the theory of gravity, but also  in our formulation of non-relativistic and relativistic quantum mechanics.
What interests us here is that even in quantum mechanics space and time are still treated as classical entities. For instance, the classical notion of space-time appears in the argument of relativistic fields; it also appears in the notion of micro-causality, in the definition of the notion of rescaling and renormalization group,  in the hypothesis of the locality of interactions, in the definition of the S-matrix  and in the hypothesis of separation of scales.
Without a preconceived classical notion of space-time, it is not even possible to define these notions, and hence what is meant by a relativistic version of quantum mechanics.
At the non-relativistic level we have on the other hand a dichotomy. Finite-dimensional quantum mechanics can be  defined purely algebraically  without any preconceived notion of classical space and the notion of time evolution can also be abstracted as a form of quantum computation \cite{NielsenChuang}. 
Schr\"odinger quantum mechanics, which can be seen as the non-relativistic precursor of relativistic field theory, on the other hand relies on a classical notion of space, appearing as the argument of the wave function, while the flow along a classical time is encoded into a choice of Schr\"odinger evolution. 
The choice of Schr\"odinger dynamics contains then a built-in notion of classical locality which we call here {\it absolute locality}: a notion of space and locality independent of the quantum probe. Of course, it is fundamental to quantum theory that a choice of basis for a Hilbert space is immaterial, and that a choice can be made subject to convenience. On the other hand, in relativistic field theory, a preconceived notion of space is built into the theory, distinguished from other bases by locality. 
If we take the point of view however, that all physics is fundamentally quantum,  it is natural to investigate whether we can infer the concept  of space and
eventually time from quantum mechanics itself.
This is the point of view we explore here. It is important to emphasize that ultimately we will not be  discussing here the quantum dynamics of a particle in a fixed space, but rather the quantum mechanics of space itself. 

Let's begin with the notion of space in non-relativistic quantum mechanics.
The basic definition that we propose for a quantum space  is that it is a choice of polarization for the representation of the Heisenberg algebra.
If one takes the Schr\"odinger representation ${\cal H} =L^2(\mathbb{R}^d)$ with wavefunctions $\vp(q)$ we recover usual space as a label for the quantum states. This representation amounts to diagonalizing the set of commuting operators which are functions of $\hat{q}$ only.
What is interesting to us  is the fact that the generic representation of the Heisenberg algebra is {\it not} of the Schr\"odinger type. As we are about to see, a generic choice of polarization corresponds to a choice of a modular cell within phase space. Hence generically,  quantum space is  modular.

We use the word {\it modular} 
because of a precise analogy with similar variables that play a natural role in purely quantum phenomena in ordinary quantum mechanics. Modular variables are described in detail by Aharonov and Rohrlich \cite{aharonov2008quantum}.  The fundamental question described there and initially posed by Aharonov  was as
follows: what type of quantum operators does one need to use in order to  capture interference effects in the Heisenberg picture?
For example, what are the quantum observables that can measure the relative  phase responsible for interference in a double-slit experiment? 
The striking answer is that no polynomial functions $P(\hat{p},\hat{q})$ of the fundamental Heisenberg 
operators, satisfying \cite{BornJordan1926}
\be\label{HA}
[\hat{q},\hat{p}]=i\hbar,
\ee
can detect such phases. Suppose that we look at a double slit experiment where the slit separation is $R$ and denote by 
$\bra \cdot \ket_\alpha $ the expectation value in a state  $\psi(x) + e^{i\theta_\alpha} \psi(x+R)$, where each $\psi$ factor is localized around one of the slits.
Here $\alpha $ is a control parameter like a magnetic flux or potential difference that modifies the interference phase.
It is easy to check that  $ \partial_\alpha \bra P\ket =0$ for {\it any} polynomial function $P(\hat{p},\hat{q})$! 
In order to detect the phase, we must work with operators such as $V(\hat p)=e^{i R \hat{p}/\hbar}$, as they do have an expectation value that depends on $\alpha$. Similarly, it is natural to introduce  $U(\hat q):=e^{2\pi i \hat{q}/R}$. 
A simple fact about  $U$ and $V$ is that, although one depends exclusively on $\hat{p}$ and the other on $\hat{q}$, they nevertheless commute \beq\label{UVcomm} UV=VU. 
\eeq
What is remarkable about this identity is that it is a purely quantum identity with no classical analog. No non-trivial function of $p$ can Poisson-commute with a non-trivial function of $q$.
Our usual preference for working with $\hat q$ and $\hat p$ is that they have classical analogues, while this is not true of $U$ and $V$. 
Since they commute they can be simultaneously diagonalized. This is not without subtlety however, as $U$ and $V$ have periodicity. The knowledge of $U,V$ therefore defines  a torus, or \hlt{modular cell}, for instance  $\mathbb{M}= [0,R)\times [0,2\pi\hbar /R)$, which is the phase space analog of the choice of a Brillouin cell in crystals.
This cell is {\it twice} the dimension of the classical polarization space determined by diagonalizing the classical variable $\hat{q}$. 

A key property of these modular operators is that they involve a length scale, which determines their periodicity. What we will find is that such scales are present in ordinary quantum mechanics in that they are associated with generic representations of the Heisenberg group. 
To explore this, let us introduce a length scale\footnote{It is of interest to note that
within the framework of purely quantum phenomena discussed extensively by Aharonov \cite{aharonov2008quantum}, a length scale (such as a slit spacing) that is inherent to the experimental apparatus is introduced within the Schr\"odinger representation. The result is a {\it contextual} modular space with an environmentally determined length scale. 
} 
$\lambda$, which in turn defines (given $\hbar$) a fundamental energy scale $\varepsilon = 2\pi \hbar c /\lambda$. Whereas $\hbar$ can be thought of as a measure of fundamental areas in phase space, the ratio $\varepsilon/\lambda$ has the interpretation of a fundamental {\it tension}, or more pragmatically, $\varepsilon$ and $\lambda$ provide scales to which we can independently measure energies and lengths. 

Given  a scale, we can unify space and momentum into phase space, denoted $\Ph$, and introduce dimensionless coordinates $x\equiv q/\lambda$ and $\tx \equiv p/\varepsilon$ measuring position and momentum in natural units. It will be convenient to use the phase space coordinatization 
\beq\label{XY}
\hat\X^{A} 
\equiv \left( \begin{array}{c} {\hat{x}^{a}} \\  \hat{\tx}_{a}\end{array} \right),\qquad 
[\hat{x}^a,\hat\tx_b ]= \frac{i }{2\pi} \delta^a{}_b .
\eeq
The Heisenberg group $\Heis_{{\cal P}}$ 
\cite{Weylbook}
  is generated by the Weyl generators, which are the phase space analogues of plane waves
\be\label{Weyl}
W_{\mathbb{K}} \equiv e^{2\pi i\, \omega(\mathbb{K},\hat{\X})},\qquad W_{\mathbb{K}}^\dagger =W_{-\mathbb{K}}.
\ee
We have introduced the symplectic structure $\omega$ given in these coordinates by \be 
\omega(\X,\Y)\equiv \tx\cdot y-x\cdot\tilde{y}.
\ee
The Weyl generators form a projective group which is a central extension of the 
translation  group on $\Ph$:
\be\label{UU}
W_{\mathbb{K}}W_{\mathbb{K}'} = e^{i\pi\, \omega(\mathbb{K},\mathbb{K}')} W_{\mathbb{K}+ \mathbb{K}'},
\qquad
W_{\mathbb{K}}W_{\mathbb{K}'} = e^{2\pi i\, \omega(\mathbb{K},\mathbb{K}')}W_{\mathbb{K}'}W_{\mathbb{K}},
\ee
where $\mathbb{K}=(k,\tilde{k}) \in \Ph$. 
 There is a natural projection $\pi:\Heis_{{\cal P}}\to \Ph$  from the Heisenberg group to the space $\Ph$, viewed as an Abelian group, given by
$\pi:W_{\K}\mapsto\K$. The first relation in (\ref{UU}) introduces a 2-cocycle defining $\Heis_{{\cal P}}$ as the central extension\footnote{ In the mathematics literature, the combination $\eta W_{\mathbb{P}}$, where $\eta \in U(1)$, is expressed as the pair $(\eta, \mathbb{P})\in U(1)\times {\cal P}$ 
which makes explicit the bundle structure $\Heis_{{\cal P}} \simeq U(1)\times {\cal P}$ and the group multiplication reads $ (\eta,\mathbb{P})\cdot (\eta',\mathbb{P}') = (\eta \eta' e^{i\pi \omega(\mathbb{P}, \mathbb{P}')}, \mathbb{P}+ \mathbb{P}')$. This is the sense in which $\Heis_{\cal P}$ is a  central extension of the translation group.} of $\Ph$. 
The closure of the algebra generated by the Weyl operators forms a $\mathbb{C}^*$-algebra and it is this algebra that we refer to as the Heisenberg algebra.

The projection $\pi:\Heis_{{\cal P}}\to \Ph$ defines a line bundle over $\cal P$. We focus on sections $\Phi$ of degree $1$ which are acted upon by the regular representation. The sections (which can also be thought of as homogeneous functions on the total space of degree one) will be referred to as states, and denoted for simplicity\footnote{A state is a homogeneous function of degree $1$ on the group which satisfies
$
\Phi(\eta\hat W_\Pm)=\eta\Phi(\hat W_\Pm):= \Phi(\Pm).
$
We can equivalently denote them as 
 functions $\Phi(\eta,\Pm)= \eta\Phi(1,\Pm)\equiv\eta\Phi(\Pm)$. The group acts on the states by \be (\eta',\Pm')\cdot\Phi(\Pm)=\Phi(\hat W_\Pm\hat W_{\Pm'} \eta')=\eta'e^{i\pi\omega(\Pm,\Pm')}\Phi(\Pm+\Pm'),\ee which yields (\ref{HeisActionOnSections}).} by $\Phi(\Pm)$. The Heisenberg group acts by right multiplication, which reads 
 \beq\label{HeisActionOnSections}
 W_{\mathbb{P}'} \Phi (\mathbb{P}) = e^{i\pi \omega(\mathbb{P}, \mathbb{P}')} \Phi(\mathbb{P}+\mathbb{P}'). 
 \eeq
 Thus the states can be thought of as living in $L^2({\cal P})$, and carry a representation of $\Heis_{\cal P}$.
 This representation is exactly self-dual: if one defines the Fourier transform 
 \be
 \Phi(\X):= \int \rd \mathbb{P}\  e^{i\pi \omega(\mathbb{P},\X)} \Phi(\mathbb{P}),  
 \ee
 we still have that $W_{\mathbb{P}} \Phi(\X) = e^{i\pi \omega(\X, \mathbb{P})} \Phi(\X+\mathbb{P})$, so this representation coincides with its  Fourier transform.

What we have described here is often regarded as ``prequantization": the  representation $L^2({\cal P})$  is highly reducible. We therefore need to restrict this highly reducible representation to an irreducible one. 
In  the geometric quantization program \cite{WoodhouseGQ}, the representation is reduced by the choice of a classical polarization, that is a Lagrangian submanifold $L\subset {\cal P}$. The states then descend to $L^2(L)$. This can always be thought of in terms of a projector acting on $L^2({\cal P})$. In the familiar case of the Schr\"odinger representation, we have a real polarization, obtained by
 \be
 P_{L} \Phi(x,\tx) := \Phi(x,0).
 \ee
In the mathematical literature it has proven more convenient to choose  complex polarizations instead. These correspond to a choice of Lagrangian subspace $P$ of the complexified phase space, and they are such that 
${\cal P} \otimes \mathbb{C} = P \oplus \bar{P}$. The resulting sections are holomorphic with respect to a complex structure $I$ on ${\cal P}$. In either of these cases, the projection is associated with a {\em classical} phase space structure. As we now describe, this is in general too restrictive: there exist additional `purely quantum' polarizations, associated with quantum Lagrangians, which we also call modular spaces. In other words the modular quantization that we propose can be seen as a generalization of the geometric quantization program to spaces that include a fundamental scale.

All of these cases have in common the choice of a commutative subalgebra of $\Heis_{\cal P}$, which is then represented diagonally. In the Schr\"odinger representation, this is the algebra of functions $O(\hat{q})$ associated with $L$, while for the complex polarizations, it is the algebra of holomorphic functions $O(\hat{z})$. The more general principle that we employ then is that an irreducible representation  diagonalizes a commutative subalgebra which is {\it maximal} in the sense that it contains its commutant.
We will also restrict to subalgebras that are closed under the $\dagger$ operation, called $*$-subalgebras, since we are interested in generalizing the notion of space itself which is a real structure. 
In summary, a quantum Euclidean space is associated with a maximal commutative $*$-subalgebra of the Heisenberg algebra.
The main point is that these are  not necessarily associated with a classical Lagrangian structure. 

In fact, this notion also fits well with familiar ideas in non-commutative geometry. In the case of commutative *-algebras, we can regard the elements of the algebra as the functions on an associated (compact) space, the Gelfand-Naimark theorem \cite{GelfandNaimark}. 
{\it We will take the point of view that any choice of a maximally commutative *-subalgebra of the Heisenberg algebra can be thought of as defining our concept of quantum Euclidean space}, this space being the Gelfand-Naimark dual of the chosen commutative algebra.


\subsection{Modular Space as Quantum Polarization}

We are now ready to apply this general strategy. Since we are interested in this work in the generalization of quantum Euclidean space we restrict our analysis to polarizations that preserve the linear structure of the Heisenberg algebra.    In other words  we need to find the maximally commuting subgroups of the Heisenberg group just defined
and construct the representation that diagonalizes this commuting subgroup.
This question was first investigated  by Mackey in a seminal work \cite{Mackey1949DukeJ}
and expanded upon in several monographs \cite{Mumford1,  PolishchukBook} . We review here these concepts well known to mathematicians but largely ignored by physicists, and present them in a new light. We have included a summary of the main results at the end of this section for the reader who wishes to skip the detailed mathematical derivations.

In the classical setting, the commutativity of observables is ensured by demanding that the symplectic potential vanishes on those observables. In the quantum case, the commutative relation involves a phase and the commutativity condition instead reads 
\be
[W_{\Pm}, W_{\Pm'}] =0\quad  \Rightarrow\quad  e^{2\pi i\omega(\Pm,\Pm')}= 1.
\ee
The maximal commutative subgroups $\hat{\Lambda}$ of the Heisenberg group are therefore easy to classify.
They correspond to {\it lattices} $\Lambda=\pi(\hat{\Lambda}) \in \Ph$ which are integral and self-dual with respect to $\omega$. In other words $\hat{\Lambda}$ is the subgroup of $\Heis_{\cal P}$  generated by elements of the form $\alpha W_\lambda$ where $\alpha \in U(1)$ and $\lambda \in \Lambda$. Let us recall that 
an integral  lattice $\Lambda$ is a subgroup of the translation group such that   $\omega(\lambda,\mu)\in\mathbb{Z}$ for all ${\lambda},\mu\in\Lambda$.
A lattice is said to be self-dual  if and only if  the condition that 
$\omega(\lambda,\mu)\in\mathbb{Z}$ for all ${\lambda}\in\Lambda$ implies that 
$\mu \in \Lambda$. The self duality condition expresses the maximality condition,  that there are no elements outside the subgroup $\hat{\Lambda}$ which commute with $\hat{\Lambda}$.

The group $\hat{\Lambda}$  is a $U(1)$ extension $\pi:\hat{\Lambda} \to \Lambda$ of the integral  lattice. A key ingredient in the construction of modular representations is the possibility to find a {\it lift} $\alpha:\Lambda\to\hat{\Lambda}$ that maps the lattice into an Abelian subgroup $\hat\Lambda_\alpha\subset\hat\Lambda$, such that $\pi\circ\alpha=id$. More precisely, we parametrize this lift in terms of a map $\alpha:\Lambda\to U(1)$ such that $ U_\lambda\equiv \alpha(\lambda)W_\lambda \in \hat\Lambda_\alpha$ is a homomorphism
\beq\label{Ucomm}
U_\lambda U_\mu =U_{\lambda+\mu}.
\eeq
This means that $\alpha(\lambda)$ must have the property
\beq\label{alphadef}
\alpha(\lambda) \alpha(\mu) e^{i\pi \omega(\lambda,\mu)}= \alpha(\lambda +\mu),\qquad \lambda,\mu\in\Lambda.
\eeq
The relation (\ref{Ucomm}) implies that the lattice is faithfully represented in the Heisenberg group.
It is similar to the property of the Duflo map which 
 is a choice of an ordering prescription for the quantization of invariant functions on a Lie group that preserves all the classical relations.\footnote{
The Duflo-map \cite{Duflo1977},  a generalization of the Harish-Chandra isomorphism \cite{HarishChandra}, is a map from the symmetric algebra Sym$(\mathfrak{g})$ over a Lie algebra $\mathfrak{g}$ to its universal enveloping algebra U$(\mathfrak{g})$. It is uniquely characterized by the
condition that  
\be
Q : \mathrm{Sym}(\mathfrak{g})^G \to  Z(U(\mathfrak{g})), 
\ee
is an isomorphism between invariant polynomials on $\mathfrak{g}^*$ and the center of U$(\mathfrak{g})$.}

An important class of lifts that we will focus on are  associated with a {\it bilagrangian structure} in ${\cal P}$, that is a choice of decomposition $\Ph= L \oplus \tilde{L}$ as a sum of two transversal Lagrangian subspaces. 
Here, we say that $L$ is Lagrangian if  $L^\perp=L$, with orthogonality  defined relative to $\omega$, and transversality means that $L\cap \tilde{L}$ is point-like.  
The choice of a bilagrangian structure can be encoded into the choice of a {\it polarization metric} $\eta$:  given the decompositions $\mathbb{P}=p+\tilde{p}$ and $\mathbb{Q}=q+\tilde{q}$ with $p\in L$, $\tilde{p} \in \tilde{L}$ etc., we define $\eta(\mathbb{P},\mathbb{Q}) = \omega(\tilde{p},q)+\omega(\tilde{q},p)$. This is  a metric of signature $(d,d)$ such that $L$ (resp. $\tilde{L}$)  are null subspaces. If one works in Darboux coordinates associated with   
the bilagrangian structure, we can simply take the symplectic structure to be the skew symmetric combination 
$\omega(\mathbb{P},\mathbb{Q})= q\cdot\tilde{p}-p\cdot\tilde{q}$ while the polarization metric is given by the symmetric combination  
\be 
\eta(\mathbb{P},\mathbb{Q})\equiv \tilde{p}\cdot q+p\cdot\tilde{q}.
\ee 
More generally given a symplectic structure $\omega$ and a bilagrangian decomposition ${\cal P}= L\oplus \tilde{L}$ we can define the neutral metric $\eta$ by the conditions
\be\label{etadef}
\eta(L,L)=0 =\eta(\tilde{L},\tilde{L}),\qquad \eta(\tilde{L},L)= \omega(\tilde{L},L),
\ee
where the notation $\eta(L,\cdot)$ means that we contract $\eta$ with any vector in the Lagrangian $L$.  This definition is tailored to make  the chosen Lagrangian subspaces to be null with respect to the polarization metric, and the non-zero pairing between conjugate Lagrangians to be given by the symplectic pairing.
The fact that the corresponding metric is neutral (i.e., of signature $(d,d)$) follows from the fact that its null subspaces are $d$-dimensional.

To construct a lift $\alpha(\lambda)$, we will make use of $\eta$. However, we must further suppose that the lattice is also compatible with the bilagrangian structure.
Given a symplectic lattice $\Lambda$ and a bilagrangian decomposition we can define sublattices $\ell =\Lambda \cap L$
and $\tilde{\ell} =\Lambda \cap \tilde{L}$ which are Lagrangian by construction. We say that the lattice $\Lambda$ is bilagrangian if the Lagrangian sublattices generate the whole:
\be
\Lambda = \ell\oplus \tilde \ell.
\ee
 This means that $\Lambda$, which is defined as a lattice self-dual with respect to $\omega$, can also be viewed as a lattice which is integral and self-dual with respect to $\eta$. Moreover the Lagrangian sublattices $\ell$ and $\tilde{\ell}$ are null with respect to $\eta$. In other words a bilagrangian lattice is a Narain lattice \cite{Polchinski:1998rq}. 
 We define \hlt{modular quantization} to correspond to the case where the lift $\alpha(\lambda)$ is constructed in terms of the polarization metric 
\beq\label{alphasol}
\alpha_\eta(\lambda):=e^{i\frac{\pi}{2}\eta(\lambda,\lambda)}.
\eeq
We can now check that this solves (\ref{alphadef}) which is equivalent to the condition 
\beq \label{comp}
\eta(\lambda,\mu)=\omega(\lambda,\mu)\quad (\mathrm{mod}\ 2) ,\quad\rm{for}\quad \lambda,\mu \in \Lambda.
\eeq
This is guaranteed if the lattice is bilagrangian as a short computation shows: 
we first notice that the defining conditions (\ref{etadef}) for the polarization 
metric imply that the following components vanish: 
$(\eta-\omega)(L,L)= (\eta-\omega)(\tilde{L},\tilde{L})=(\eta-\omega)(\tilde{L},L)=0 $. So for general $\lambda=(n,\tilde{n}) \in \Lambda $ and $\mu=(m,\tilde{m}) \in \Lambda $ we have that 
\[ (\eta-\omega)(\lambda,\mu)
= 2\omega(n,\tilde{m}),\] where we used the fact that $\ell$ and $\tilde{\ell}$ are Lagrangian. The righthand side is zero (mod 2) because $\omega(x,\ty)\in\mathbb{Z}$ for $\lambda,\mu\in\ell\oplus\tilde\ell$. 

Now that we have chosen a bilagrangian lift $\alpha_\eta$ we can look for the general solution of the lift condition (\ref{alphadef}). It is clear from this condition that given any lift $\alpha$ the combination $\alpha/\alpha_\eta$ is an abelian character. It is therefore uniquely characterized by the choice of 
a point $\mathbb{O} \in \Ph/\Lambda $ inside the modular cell.\footnote{We have  denoted the characteristic as $\mathbb{O}$ since it corresponds to a choice of observer inside the cell, as we will see.} We call this point the {\it characteristic} of the modular cell.
This shows that a general lift can therefore be parametrized by a bilagrangian structure encoded into $\eta$ and a characteristic $\mathbb{O}$ and is given by 
\be 
\alpha_{(\eta,\mathbb{O})}(\lambda) = e^{i\frac{\pi}{2}\eta(\lambda,\lambda)} e^{ 2 \pi i\omega(\mathbb{O},\lambda)}.
\ee
This shows in particular that we can always understand a choice of another modular lift associated with a different polarization metric $\eta'$ as being related to a lift with characteristic. In other words, given two polarization metrics $\eta$ and $\eta'$  compatible with $\omega$ in the sense (\ref{comp}), we can always find a characteristic $\mathbb{O}_{\eta,\eta'}$ such that 
\be
\alpha_{\eta'} = \alpha_{(\eta, \mathbb{O}_{\eta,\eta'})} .
\ee
As a summary, we have seen that the choice of modular quantization $\alpha_\eta$ is associated with the choice of a polarization  metric $\eta$ which determines a bilagrangian lattice $(\Lambda,\eta)$, also called Narain lattices.
We have seen that the identification of  the  symplectic lattice $(\Lambda,\omega)$ as a Narain lattice $(\Lambda,\eta)$ allows us to  lift $\Lambda$   into a commutative subgroup of the Heisenberg group. This correspondence between symplectic lifts and Narain lattices is not fortuitous, and goes a long way in explaining why modular spaces naturally appear in string theory. More generally, they should be regarded as relevant in any proper quantum theory of gravity, that is any quantum theory with a fundamental scale built in.

%

From now on we restrict for illustration purposes to lifts of zero characteristic, that is to modular quantizations.
Given the polarization metric, we can take the associated lift morphisms to be the restriction to $\Lambda$ of the maps
\beq
\alpha_\eta(\mathbb{K}) \equiv e^{\frac{i\pi}{2} \eta(\mathbb{K},\mathbb{K})},
\eeq
and we define the modified Heisenberg generators
\beq 
U_{\mathbb{K}} = e^{\frac{i\pi}{2} \eta(\mathbb{K},\mathbb{K})} W_{\mathbb{K}}.
\eeq
These satisfy the algebra 
\beq
U_{\mathbb{K}} U_{\mathbb{K}'} = e^{i\pi (\omega-\eta) (\mathbb{K},\mathbb{K}')} U_{\mathbb{K}+\mathbb{K}'}.
\eeq
This can be regarded as specifying a particular operator ordering, one respecting, as does (\ref{Ucomm}), the lattice translation group. 
Indeed, in the coordinates described above, we have 
for $\mathbb{K}=(k,\tilde{k})$ that \[U_{\mathbb{K}} = e^{2\pi i\, \tilde{k} \hat{x} }  e^{- 2\pi i\, k \hat{\tx} },\] where the $\hat{x}$ are moved to the left of the $\hat{\tx}$. 
We can now see that including the characteristic $\mathbb{O}$ amounts to a phase redefinition of the morphism $U_\lambda \to e^{2\pi i \omega(\mathbb{O},\lambda)} U_\lambda$ and such a phase redefinition corresponds to a choice of origin 
$\X \to \X-\mathbb{O}$. In other words, and as we will see in more detail later, the dependence of the lift on the characteristic is what allows the preservation of translation invariance.

Given the bilagrangian lift we can now define the corresponding Hilbert space $ {\cal{H}}_\Lambda $.
The idea behind the construction is simple and goes back to 
Mackey \cite{Mackey1949DukeJ, Mackey1949PNAS, Mackey1952, Mackey1958}.
We have seen that a quantum polarization is a choice of a maximal commutative subalgebra $\mathbb{C}[\hat\Lambda_\alpha]$. Here $\hat{\Lambda}_\alpha \to \Lambda$, with U$(1)$ fiber, is the Abelian subgroup lifted by $\alpha$ and associated with $\Lambda$. As we have seen, the quantum algebra acts naturally on degree $1$ sections, that is, homogeneous functions on the group $\hat{\Phi}(\eta W_{\Pm})= \eta \hat{\Phi}( W_{\Pm})$.
Henceforth, it will be convenient to change notation slightly and denote the section by $\varphi(\X):=\hat{\Phi}(U_\X)$.

It is important to remark that because of the non-commutative structure we in fact have two different actions of the group on the algebra, one from the left and one from the right.  The two actions are defined by 
\be
L_{U_\Y} \varphi(\X) = \hat{\Phi}(U_\Y^{-1} U_\X ),\qquad 
R_{U_\Y} \varphi(\X) = \hat{\Phi}(U_\X U_\Y).
\ee
We require that the left-action of the lifted lattice $\hat{\Lambda}$ be trivial,
$
L_{U_\lambda}\varphi(\X)=\varphi(\X),
$
which implies that
\beq\label{leftinv}
\varphi(\X+\lambda)=e^{i\pi(\eta-\omega)(\lambda,\X)}\varphi(\X),
\eeq
and then it follows that the right-action of $\hat{\Lambda}$  is diagonalized
\beqn\label{QP1}
R_{U_\lambda}\varphi(\X)=e^{-i\pi(\omega+\eta)(\lambda,\X)}\varphi(\X+\lambda)=e^{-2\pi i\omega(\lambda,\X)}\varphi(\X).
\eeqn
Given the form of (\ref{leftinv}), we identify $L_\Lambda \to T_\Lambda$ to be the $U(1)$ bundle over $T_\Lambda={\cal P}/\Lambda$ given by the identification 
\beq
(u,\X) \sim(ue^{i\pi (\omega-\eta)(\lambda,\X)},\X+\lambda),
\eeq
and $\varphi$ with $L^2$ sections of this bundle. 
Such sections are quasi-periodic on the torus. Note however that the modulus of the field is periodic and therefore the $L^2$-norm 
$
 ||\vp||^2 = \int_{\Ph/\Lambda} | {\vp(\X)}|^2 \rd \X
$
 is well-defined as an integral over the torus.

To summarize: we have seen in this section that a typical commutative sub-algebra of the Heisenberg algebra  is given by {\it modular} observables which are functions on $\Ph$ periodic with respect to a lattice $\Lambda \subset \Ph$.
Representations that diagonalize these modular observables require the introduction of a bilagrangian structure, that is, the choice of a decomposition 
of $\Ph =L\oplus \tilde{L}$, where $(L,\tilde{L})$ are Lagrangian subspaces, which extends to the lattice $\Lambda =\ell\oplus \tilde{\ell}$.
This bilagrangian structure can be equivalently characterized by a choice of a neutral metric $\eta$ for which $L$ and $\tilde{L}$ are null and such that $\eta(\tilde{L}, L) =\omega(\tilde{L},L)$.
Given these structures, the Hilbert space ${\cal H}_{(\Lambda,\eta)}$ is given by $L^2$  sections of a line bundle over a $2d$-dimensional torus $T_\Lambda= \Ph/\Lambda$ in phase space . In other words, states are {\it quasi-periodic} functions with the quasi-periods  determined by the bilagrangian structure:
\be
\varphi(\X+\lambda) = e^{i\pi(\eta-\omega)(\lambda,\X)}\varphi(\X),
\ee
and the action of the Heisenberg group element $U_\PP=  e^{\frac{i\pi}{2}\eta(\PP,\PP)} e^{2i\pi \omega(\PP,\hat{\X})}$ is given by 
\be
U_\PP \, \varphi(\X) = e^{-i\pi (\eta +\omega)(\PP,\X)}\varphi(\X+\PP),
\ee
on $T^d$ with $T= S^1\times S^1$, instead of functions over $\mathbb{R}^d$.


\subsection{Modular space as Fundamental Space}

We  draw  several conclusions from the above analysis.
It is first important to appreciate that although the commutative observables $U_\lambda(\hat{\X})$  are $\Lambda$-periodic, i.e., $U_\lambda(\hat{\X}+\mu)=U_\lambda(\hat{\X})$ as operators,  the corresponding modular fields $\vp(\X)$ that diagonalize them are not, they are only quasi-periodic, as in eq. (\ref{leftinv}). This means that we should think of the modular field as living on a modular cell $\mathbb{M}$ in phase space, which is the analogue of a Brillouin cell. More precisely a modular cell $\mathbb{M}$ corresponds to a choice of section of a line bundle over  $ T_\Lambda = \Ph/\Lambda$. The quasi-periodicity (\ref{QP1})   ensures that the knowledge of $\vp$ on $\mathbb{M}$ determines the knowledge of $\vp$ on all phase space. Once a modular cell is chosen, we can then describe the modular Hilbert space as ${\cal H}_\Lambda = L^2( \mathbb{M}_\Lambda)$.
Now it is true that as a Hilbert space, we have $L^2( \mathbb{M}_\Lambda)\simeq L^2(T_\Lambda)$. This isomorphism depends however on the choice of modular cell, while the original formulation does not. It also does not respect the continuity properties since it ignores the quasi-periodicity condition. For instance, it is well-known that a differentiable quasi-periodic function $\Phi$ necessarily vanishes inside each cell while a periodic function does not have to. This follows from the fact that 
the quasi-periodicity translates into the condition that the winding number of $\Phi$ is $1$:
\be
\frac1{2\pi i} \oint \frac{\rd \Phi}{\Phi} =1,
\ee
where the integral is along a cycle that winds once around the boundary of the modular cell. 
 For these reasons, we see that it is more 
appropriate to think of the argument of the wave function to be a modular cell 
and not a torus. This distinction is somewhat subtle, but important conceptually.

These representations differ drastically from the Schr\"odinger representation in three respects. First, the spectrum of a modular observable defined as a modular space, is compact, and forms a torus. Second, the dimension of such a modular space is doubled as compared to the Schr\"odinger polarization. In other words, states are sections  over
$T^d$ with $T= S^1\times S^1$ instead of functions over $\mathbb{R}^d$.
Third, modular space is not simply connected, while Schr\"odinger space is, generically, simply connected.
The fact that modular space is not simply connected allows for Aharonov-Bohm phases to be  canonically defined as quasi-periods.

\subsection{Stone-von Neumann Theorem}

In the previous section, we have seen that a generic representation of the Heisenberg algebra is modular, i.e., associated to a discrete lattice $\Lambda \subset\Ph$.
To fix ideas, let us take standard coordinates $(x,\tx)$ in ${\cal P}$ and assume that we start with the self-dual lattice given by
$\Lambda_1 \equiv \mathbb{Z}^d \times \mathbb{Z}^d$.
Other lattices can be obtained from $\Lambda_1$ by the action of a symplectic transformation and  inequivalent lattices are classified by the  quotient $ Sp(2d,\mathbb{R})/Sp(2d,\mathbb{Z})$. Let's consider for instance the symplectic transformation $(x,\tx)\to (a x,a^{-1} \tx)$, which  deforms the self-dual lattice, the modular cell and the Hilbert space into  $\Lambda_a$, $\mathbb{M}_a=[0,a]\times [0,a^{-1}]$ and ${\cal H}_a$ respectively. When $a>1$, it squeezes the modular cell  along the momentum direction and stretches it in the  space direction. In the limit  $a \to \infty$ the lattice is fully stretched in the space direction and  becomes very dense in the momentum direction so that it limits to  a classical Lagrangian $\Lambda \to \tilde{L}$, the torus $T_\Lambda=P/\Lambda$ degenerates into the quotient space $ P/\tilde{L} \simeq L$ and the Hilbert space based on the modular cell becomes the space of functions that do not depend on $p$, that is $L^2(\mathbb{M}_a) \to  L^2(L)$.
Since $L^2(L)=L^2(\mathbb{R}^d)$, we recover the usual Schr\"odinger representation in this singular limit, where the modular cell is fully squeezed.
In this limit we lose all the information about the phase space connection and the Aharanov phases. In this limit we lose the information that the cell has unit Planck area. In this limit we also lose half the dimensions of the modular cell.
From this perspective, it is clear that the Schr\"odinger representation is not a generic representation of the Heisenberg algebra but on the other hand a highly singular one.
It is only for this singular choice of lattice that one recovers the usual classical notion of space. 

Modular space has applications  in ordinary quantum mechanical settings, in which the length scale is set by an experimental apparatus (such as a slit spacing). From this point of view, what we have done above is to define the underlying theoretical structure of Aharonov's modular variables. 
On the other hand, if a {\em fundamental} scale was available, we could use the modular space as a new and fundamental notion of space.

The concept of {\em locality} that we use in physical theories is tied to the choice of a preferred basis, the Schr\"odinger polarization. As we have just seen this preferred basis is from the point of view of quantum mechanics very singular while the modular basis is generic. 
If our usual notion of space is singular, it is of interest to consider why we
use it as the foundation on which the rest of physics is built. We can identify three reasons for this. The first is historical: the notion of space came before quantum mechanics (Euclid lived 2300 years ago) and it was natural to build in classical Euclidean and Minkowskian geometries into quantum mechanics instead of the other way around.
The second reason is that, even if classical space arises 
as a limit or approximation of modular space, it is a very robust approximation. 
Indeed, if we identify the scale associated with the modular cell with the Planck scale, the squeezing of the modular cell is given by the {\it gravitational tension} $\varepsilon/\lambda \sim c^4/G_N \simeq 10^{17} kg / \AA$.
This is a huge tension and all quantum experiments performed today,
even the most energetic ones,  fail by a huge margin to probe the thickness of the modular cell in the momentum direction. For all practical purposes we can then assume that the space is not modular, as long as we are not dealing with quantum gravity observables. 
The third reason is mathematical and ingrained deeply into the psyche of any quantum physicist. It comes from the Stone-von Neumann theorem \cite{Stone1930, vonNeumann1931}, which states that all representations of the Heisenberg group are {\it unitarily equivalent} to the Schr\"odinger representation. 
It is this theorem that seems to justify the use of a classical space  as a universal polarization. 
This very powerful mathematical theorem is in our view fundamentally  misleading in the present context.
It doesn't appreciate that different polarizations do have very different geometry, notions of regularity, and sets of $C^\infty$ vectors. 
In particular, the spectrum of operators diagonalized by the 1d Schr\"odinger representation is non-compact and 1-dimensional while the modular representation is compact and 2-dimensional. 

In fact the original proof \cite{vonNeumann1931} by von Neumann of the theorem is non-constructive. It uses the fact that the operator $P \equiv \int e^{-\pi(x^2 +\tx^2 )} W_{(x,\tx)}$ is a rank one projector on any irreducible representation and the unitary equivalence ${\cal H}\simeq {\cal H}'$  can simply be implemented by identifying the image  $P{\cal H}= P {\cal H}'$.
This projector crushes all the information about the representation and any 
choice of representation is valid. From that perspective it is not clear why one should focus on the Schr\"odinger representation more than any other, except that this was the only one explicitly known at the time of the theorem.

 The explicit construction by Mackey \cite{Mackey1949DukeJ, Mackey1949PNAS, Mackey1952, Mackey1958} twenty years later of induced representations that we presented in the previous section, gives on the other hand a wealth of explicit representations based on modular spaces. Any one of these is a valid polarization and we just argued that the self-dual polarization is more typical and less singular than the Schr\"odinger one in the sense that all modular spaces can be simply obtained by a symplectic mapping from the self-dual one, while the reverse is not true --- the modular spaces cannot be understood in terms of the usual Schr\"odinger space. Our goal is now to rebuild the notion of dynamics, the relativity principle and fields in the context of modular spaces.
 It is interesting to note that doing so amounts to going back to the original formulation of quantum mechanics invented by Heisenberg, Born, Jordan and Dirac \cite{BornJordan1926, BornHeisenbergJordan1926, Dirac1925, Dirac1926}. Their discovery of quantum mechanics was intimately linked with a deconstruction of the notion of space and this was one of the most puzzling aspects of matrix quantum mechanics. This was the state of the art before Schr\"odinger and then von Neumann reintroduced into quantum mechanics the usual notion of space that matrix quantum mechanics had deconstructed.


\subsection{Zak transform }

The Stone-von Neumann theorem implies that there is an isomorphism between the modular and Schr\"odinger representations.
In order to illustrate this unitary equivalence we will restrict ourselves to $d=1$ to simplify the exposition and we choose the canonical symplectic structure and commutation relation 
$[\hat{q},\hat{p}]=i\hbar$  which are mapped onto 
the dimensionless parameters $x\equiv q/\lambda$ and $\tx \equiv p/\varepsilon$, with $\lambda \varepsilon=2\pi \hbar$, measuring position and momentum in natural units. 
We restrict ourselves in this section to the  self-dual lattice $\Lambda_1$ with elements denoted $\lambda=(n,\tn) \in \mathbb{Z}^2$. Any other lattice in 1d is related to this one by rescaling $(x,\tx)\to (a x,a^{-1}\tx)$, which corresponds to a different choice of fundamental scales preserving $\hbar$.

Bringing to this example the general discussion of the previous section, we find that the self-dual modular space is represented by the two-dimensional cell $\mathbb{M}_1 = [0,1)\times [0,1)$ and the Hilbert space 
${\cal{H}}_1$ is given by the space of  $L^2$ quasi-periodic functions  (\ref{leftinv}). We can write the quasi periodicity condition explicitly in terms of the Lagrangian coordinates $(x,\tx)$ as 
\be\label{QP2}
\Phi(x+n,\tx+\tn) = e^{2i\pi n\tx} \Phi(x,\tx),\qquad |\!|\Phi|\!|^2 = \int_{[0,1)^2}  \rd x \rd \tx
|\Phi|^2  < \infty.
\ee
This Hilbert space can also be understood as the space of 
$L^2$ sections of a $U(1)$ bundle over the torus $T_2 = \mathbb{R}^2/\mathbb{Z}^2$. 
The Zak 
transform\footnote{Apparently, initially discovered by Gelfand, the Zak transform \cite{Zak1967, Zak1968} is an important concept, for example, in signal processing \cite{SamplingTeory}. It also makes an appearance in condensed matter contexts such as the Quantum Hall Effect \cite{Thouless1982}.} is an explicit unitary map $\vp(q)\to (Z_\lambda\vp)(x,\tx)$ between  the Schr\"odinger Hilbert space and the modular one
\be
Z_\lambda: L^2(\mathbb{R}) \to L^2(\mathbb{M}_1
), \qquad 
(Z_\lambda\vp)(x,\tx) \equiv \sqrt{\lambda} \sum_{n\in \mathbb{Z}} e^{-2\pi in \tilde x} \vp(\lambda(x+n)).
\ee
It can be checked that the image satisfies the quasi-periodicity property and that  it is a unitary map.
This equivalence is non-trivial from a topological perspective: it maps functions on a non-compact space onto functions on a compact space of twice the dimension. The fact that such an isomorphism is possible at the quantum level is the expression of the complementarity principle.
The inverse of the Zak transform  reads \bea\label{ZF}
(Z_\lambda^{-1}\Phi) (x +n) &=&
\frac{1}{\sqrt{\lambda}}\int_0^1\rd \tx\ e^{2i\pi n \tx} \Phi(\lambda^{-1}x,\tx), \quad x \in [0,1).
\eea
We see from this expression that while it is an $L^2$ isomorphism the inverse map does not necessarily respect continuity properties.  The image under $Z^{-1}_\lambda$  of a smooth modular state generically exhibits discontinuities at $x=n$ and is not in the domain of the translation operator $\hat{p}$.
The  behavior of the Zak transform under the Fourier transformation 
  $\sqrt{2\pi\hbar} \, \tilde{\vp}(p) \equiv \int \rd q \ e^{-i pq/\hbar} \vp(q)$  is 
  given by 
\bea
(Z_\lambda \tilde{\vp})(x,\tx) &=& e^{2\pi ix\tx } (Z_\varepsilon\vp)(-\tx,x).
\eea
We see that the Fourier transform acts very simply under the Zak transform; it is given, up to a phase, by the exchange $\X \to I\X$ where $I(x,\tx) =(-\tx,x)$.
It is interesting to note that in the case where $x=\tx=0$, this Fourier identity reduces to the Poisson resummation formula.

It is useful to express the form of the elementary Heisenberg  operators  under the Zak transform:
\be\label{Zakrep}
\hat{\tilde{x}}\to -\frac{i}{2\pi} \pa_x ,\qquad 
\hat{x}\to \left(\frac{i}{2\pi} \pa_{\tx} + x \right).
\ee
An important point is the fact that we could define another isomorphism of the form $Z_\lambda^{(\alpha)} = e^{2i\pi\alpha} Z_\lambda$, where $\alpha(x,\tx)$ is a function on $\mathbb{M}_1$. The image still satisfies the same quasi-periodicity conditions as long as $\alpha$ is periodic. 
But under this gauge rescaling, we get a different representation of the Heisenberg generators. In order to express the change we introduce an Abelian phase space connection $ \mathbb{A}_A \rd \X^A =A_x \rd x + A_{\tx} \rd\tx $ and the corresponding covariant derivatives $\nabla_{x} = \pa_x -{2\pi}i A_x$ and 
$\nabla_{\tx} = \pa_{\tx} -{2\pi}i A_{\tx}$. 
The initial choice of gauge for $Z_\lambda$ is $(A_x,A_{\tx})= (0, x)$. This connection is discontinuous on the torus, and it is this discontinuity which necessitates the fields to be quasi-periodic instead of periodic. In other words,
the naive derivative does not respect the quasi-periodicity relation but the covariant derivative does. We can check that indeed we have
\be
[\nabla_{\tx} \Phi](x+n,\tx+\tn)]= 
[\pa_{\tx}- 2i\pi(x+n) ][e^{2i\pi n\tx}  \nabla_{\tx} \Phi](x,\tx)]
=e^{2i\pi n\tx} [\nabla_{\tx}\Phi](x,\tx).
\ee
Under the gauge rescaling, the connection transforms as $ A\to A +\rd \alpha$ and the  canonical  generators can be expressed as covariant derivatives
\be\label{conn0}
\hat{\tilde{x}} \Phi = -\frac{i}{2\pi} \nabla_{x}\Phi ,\qquad 
\hat{x}\Phi = \frac{i}{2\pi} {\nabla}_{\tx} \Phi.\qquad 
\ee
The  Zak transform $Z_\lambda^{(\alpha)}$ corresponds to the choice $A_x=\pa_x \alpha$, $A_{\tx}= x +\pa_{\tx}\alpha$. 
This means that the curvature of the phase space connection associated with the Zak transform is constant
\be
F(\mathbb{A})= \pa_x A_{\tx} -\pa_{\tx}A_x = 1,
\ee
which is a gauge invariant statement.
In order to write in a convenient manner the quasi-periodicity conditions (\ref{QP2}),
we introduce another connection $\tilde{\mathbb{A}}$ which is given by 
 $ \mathbb{A}+ \tilde{\mathbb{A}}  =\rd (x\tx)$.
 This connection is such that it has the opposite constant curvature $\tilde{F}=-1$ and its covariant derivative $\tilde{\nabla}\equiv \pa -2i\pi \tilde{A}$ commutes with $\nabla$
 \be
 [\tilde{\nabla}_A ,\nabla_B]=0.
 \ee
The relevance of this connection appears when we rewrite the quasi-periodicity conditions (\ref{QP1},\ref{QP2}). It allows us to write the
 left-invariance condition (\ref{leftinv}) which defines the line bundle in a covariant manner as the condition
\be
\phi(\X +\lambda) = \exp\left({2\pi i\int_{(\X,\lambda)}\tilde{\mathbb{A}} }\right)\phi(\X),
\ee 
where $\lambda \in \Lambda$ and $(\X,\lambda)$ denotes a path in $\Ph$ that starts at $\X$, ends at $\X +\lambda$ and moves along the lattice $\Lambda$. The connection is not flat but if we choose two different paths, the expression differs by a  phase $ \exp(2\pi i\oint_\Lambda \tilde{\mathbb{A}})$ which is the integral along a closed path in $\Lambda$. The constancy of the curvature implies that each lattice cell carries an integer unit of flux and thus the phase is trivial. 
The quasi-periodicity condition can also be written as 
\be
e^{\lambda^A \tilde{\nabla}_{A}}  \phi(\X)
= e^{n \tilde{\nabla}_{\tx}} e^{\tn \tilde{\nabla}_x}\phi(\X)=\phi(\X),
\ee
for $\lambda=(n,\tilde{n})$.

This shows that we can understand a modular space as a phase space torus equipped with a constant `magnetic field.'  This magnetic field is not stretched in space but instead as one component along space and one along momenta.

In order to illustrate the behavior of the Zak transform, it is useful to consider the 
transform of the Gaussian state $e_{(\tau,\lambda)}(q)=\frac{1}{\sqrt{\lambda}}e^{i\pi \tau {q^2}/{\lambda^2} } $, with $\tau $ a complex parameter with Im$(\tau)>0$, controlling the shape of the Gaussian relative to the units $(\lambda,\varepsilon)$. Thus
$\tau$ provides a dimensionless squeezing parameter.
The Zak transform of the Gaussian is given, up to an overall phase, by the Jacobi theta function\footnote{\[\theta(z;\tau)\equiv \sum_{n\in\mathbb{Z}}e^{i\pi n^2\tau+2\pi iz}.\]}
\be
\Theta_\tau(x,\tx)\equiv (Z_\lambda e_{(\tau,\lambda)})(x,\tx) = e^{i\pi \tau x^2 } \theta(z;\tau),\qquad z\equiv \tau x -\tx.
\ee
The quasi-periodicity condition translates into the theta function quasi-periodicity
\be
\theta(z+\tau a +\tilde{a};\tau) = e^{-i\pi(\tau a^2 + 2 a z)}\theta(z;\tau),\qquad (a,\tilde{a})\in \mathbb{Z}^2,
\ee
and since $\tilde{e}_{(\tau,\lambda)} =\sqrt{ -{i}{\tau}}\ e_{(-1/\tau,\varepsilon)}$, the exchange property  (\ref{ZF}) of the Zak transform under the Fourier transform is equivalent to the inversion identity
\be
 \theta (z/\tau;-1/\tau) = 
\sqrt{-i\tau}\ e^{i\pi z^2/\tau}\theta(z;\tau).
\ee
In fact, we can characterize the Jacobi theta function  as the {\it unique} element  $\Theta_\tau \in {\cal H}_\Lambda$  such that
$ e^{-i\pi \tau x^2 } \Theta_\tau(x,\tx)$ is holomorphic with respect to $z=\tau x-\tx$.
The fact that there are no other quasi-holomorphic sections will be important for us.
What is special about the section $\Theta_\tau$, apart from its holomorphicity property, is the fact that it vanishes only at one point inside the modular cell, the center: $\Theta_\tau(\tfrac12,\tfrac12)=0$. 
We have already emphasized that a differentiable section needs to have at least one zero inside the cell. The theta section and its translations minimize the number of zeroes inside the cell.

\subsection{Modular Translations}

In this section we investigate the notion of ``pure space'' or empty space and whether this notion, which is essentially classical, can be generalized to modular space. In other words, we want to investigate whether there is such a thing as empty modular space? The answer, as we are going to see, is negative and this has many far-reaching consequences.
But first, let's try to formalize more what is usually meant by empty space; why it is a central concept, and how it is consistent with the Schr\"odinger representation.
The notion of empty space  is a key notion for our understanding of physics, it allows us to organize our representation of nature by imagining space as a receptacle in which we put matter. So what is usually meant by this concept of empty space?
At the classical level and in the absence of gravity what is meant is a place where no change takes place. In mathematical terms empty space has the property of being isotropic, that is, translation invariant. At the quantum level 
the notion of empty space translates into the choice of a translation invariant state, that is, a {\it vacuum} state solution of 
\be
\hat{p}_a  |0\ket=0.
\ee
Now we see that in order to define the notion of vacuum representing pure space, we need to identify a notion of translation algebra. In the Schr\"odinger picture this translation algebra is associated with the choice of a classical Lagrangian, the subset of phase space generated by $\hat{p}_a$. The question that is therefore going to occupy us is whether or not we can identify a 
notion of translation algebra for modular space.

That is, instead of  identifying the vacuum with a state that annihilates the action of a commutative algebra like the translations, we can also define the vacuum to be the minimal energy state associated with a positive operator that for brevity we refer to as a ``Hamiltonian."
Hopefully, for the vacuum state corresponding to empty space, these two notions are the same. Indeed, demanding that $|0\ket$ is annihilated by the translation generators $\hat{p}_a$ is equivalent 
to demanding that it is annihilated by the operator\footnote{Although we are using the term ``Hamiltonian" here, the reader should not confuse it with the Hamiltonian of some particle system. Instead, it is a tool that we use to identify a ``vacuum state of empty modular space."} 
\be
\hat H_h\equiv h^{ab} \hat{p}_a \hat{p}_b,  
\ee
as long as $h^{ab}$ is a positive definite metric.
The fact that this notion of vacuum state is the same as the notion of translation invariant state follows from the fact that if $\hat H_h|0\ket=0$ 
then $\hat H_{h'}|0\ket=0$ for any other positive definite metric.
In other words the vacuum state of usual flat space does not depend on the choice of metric the space possesses, only the excited states do.
This is the key property that allows one to think of empty space as a universal notion.

Now this discussion relies strongly on the fact that we are in the  Schr\"odinger representation where the algebra of translations is clearly identified, but what about modular space?
We have seen in the previous sections that the modular representations ${\cal{H}}_{\Lambda}(\alpha)$ allows us to diagonalize the modular observables  $\Phi(\hat{\X}+{\lambda})=\Phi(\X)$ generated by the lattice observables $U_\lambda$ with $\lambda \in \Lambda$. What we want to investigate now is the possibility of defining Hermitian operators $\hat{\mathbb{P}}_A$ conjugate to  $\hat{\X}^A$. 
This means that we are looking for 
operators such that 
\be\label{Comm1}
 [\hat{\mathbb{P}}_A, \Phi]= \frac{i}{2\pi}\partial_A \Phi ,\qquad \Phi(\hat{\X}+\lambda)=\Phi(\hat\X).
 \ee
Since the modular observables are angles, we expect each $\hat{\mathbb{P}}_A$ to have a purely discrete spectrum and indeed the previous commutation relations implies that 
 \be\label{Comm2}
 U_{\lambda}^{-1} \hat{\mathbb{P}}_B U_{\lambda}
 = \hat{\mathbb{P}}_B + \lambda^A \omega_{AB}.
 \ee
 It is easy to see that the most general solution of (\ref{Comm1}) is given by 
 \be
 \hat{\mathbb{P}}_A={\omega_{AB}}  \hat\X^B + a_A(\hat\X),
 \ee
 where $a_A(\hat\X)$ is an operator that commutes with all modular functions
 and is therefore a modular function itself.
 We have seen in the previous section  that $\hat{\X}$ is acting as a covariant derivative $\nabla_A^0=\pa_A + 2i\pi A_A$. Therefore this definition shows that when acting on modular states $\varphi \in {\cal{H}}_{\Lambda}(\alpha)$, $\hat{\mathbb{P}}$ acts also as a covariant derivative
 \be
  \hat{\mathbb{P}}_A \varphi(\X) = \frac{i}{2\pi}\nabla_A \varphi(\X),
 \ee
 where $ \nabla_A= \pa_A + 2\pi i(A_A+a_A) $ is a $U(1)$ connection, which is the translation of the naive phase space connection $A_A$ by $a_A$.
 The curvature of this connection is 
 \be
 [\nabla_A,\nabla_B]= 2\pi i F_{AB},\qquad F_{AB}= \omega_{AB} + \pa_A a_B -\pa_B a_A.
 \ee
 This algebra is invariant under the gauge transformation $\Phi(\X) \to e^{2i\pi \alpha(\X)}\Phi(\X)$  which implies that 
\be
\hat{\mathbb{P}}\to \hat{\mathbb{P}},\qquad a_A \to a_A +\pa_A \alpha,
\ee
where $\alpha$ is a modular function.
 The question that we need to focus on is  what choice of connection represents the translation operation associated with pure space?

 A 
 fundamental property  of  modular translations is the fact that 
 there is no choice of connection $a_A$ that can make the  translations 
 commute with each other. This follows from the fact that for any choice of connection and any closed and non-contractible 2-surface $S$ embedded in $T_\Lambda$  we have that 
 \be
 \int_S  F_{AB} \rd \X^A\wedge \rd \X^B =  \int_S  \omega_{AB} \rd \X^A \wedge \rd \X^B \neq 0.
 \ee
 This equation tells us that the ``modular  flux" through 2d modular cells in phase space, which is an analogue of the magnetic flux in a quantum Hall sample,  does not vanish but is always equal to its area. 
 This is a fundamental difference with the Schr\"odinger case where a general translation operator\footnote{That is, an operator $N_a$ such that $[N_a ,\phi(\hat{q})]=-i\pa_a \phi $.}  is of the form $\hat{p}_a + A_a(\hat{q})$, and we can choose $A_a$ to have vanishing curvature. In this case the $p_a$ commute and it makes sense to impose that they are simultaneously diagonalized.
 The modular translations on the other hand {\it never commute}.
 We cannot therefore define the state representing pure space as a state of isotropy. In this sense the study of modular space is similar to the study of the dynamics of charged particles in a non-zero magnetic field. 
 
Another way to understand this impossibility is to look at the vacuum wave-functional   
\be 
\bra \X | 0 \ket\equiv e(\X).
\ee
 The vacuum wave-functional $e(\X)$ has to be a section of the line bundle. That is, if one focuses on the one-dimensional modular space, it has to satisfy, in the Zak gauge, the quasi-periodicity conditions 
 \be
 e(x+n,\tx +\tn) = e^{2i\pi n \tx} e(x,\tx).
 \ee
 This section labels the possible vacuum states $|0\ket$. In the Schr\"odinger representation, we fix the ambiguity by demanding the vacuum to be translation invariant, which implies that $e$ is constant. 
 Since the line bundle is non-trivial, we cannot simply take $e$ to be constant.
 This follows from the fact that the quasi-periodicity condition implies the circulation of the phase of $e$ around a cycle lying in the boundary of the modular cell carries a unit of flux
\be\label{res}
1=\frac1{2\pi i}\oint \frac{\rd e}{ e }.
\ee
This is due to the fact that the phase $\vp$ given by $e(\X)\equiv |e(\X)| \exp(2\pi i\vp(\X))$ necessarily possesses a unit winding number.
The integral (\ref{res}) computes a residue and it implies that if differentiable, which we assume,  the vacuum wave function  necessarily possesses a zero.
The location of this zero, denoted $\mathbb{O}$, clearly breaks the translation invariance by defining a preferred location inside the modular cell.
In other words the fact that there is {\it no translation invariant vacuum} because the translation generators do not commute  is related to the fact that $e(\X)$ cannot be chosen to be constant and vanishes  at at least one point in the cell.

We therefore need to find another criterion in order to choose the plane wave section.
Translation invariance would be the condition $\hat{\mathbb{P}}|0\ket=0$. Since this is not possible the next natural choice, is to minimize the translational energy. 
This means that we pick a positive definite metric $H_{AB}$ on $\Ph$, and we define 
\be
\hat{E}_H\equiv H^{AB} \hat{\mathbb{P}}_A\hat{\mathbb{P}}_B,
\ee  
and demand that $|0\ket_H$ be the ground state of $\hat{E}_H$.
This is indeed the most natural choice and it shows that we cannot fully disentangle the kinematics (i.e., the definition of translation generators) from the dynamics.
In the Schr\"odinger case, since the translation generators commute, the vacuum state $\hat{E}|0\ket=0$ is also the translation invariant state and it carries no memory of the metric $H$ needed to define the energy.
In our context, due to the non-commutativity of translations, the operators $\hat{E}_H$ and $\hat{E}_{H'}$ do not commute. As a result the vacuum state depends on $H$, in other words $|0\ket_H\neq |0\ket_{H'}$, and it also possesses a non-vanishing zero point energy.

\subsection{Modular Vacua}\label{modv}

In order to understand how the vacuum state changes under a change of ``Hamiltonian" we now make the extra assumption that the metric entering the ``Hamiltonian"  is derived from a complex structure. That is we assume that $ H(\X,\Y) = \omega(\X,I(\Y))$ where\footnote{ A complex structure is usually a map $I:T{\Ph}\to T{\Ph}$ but since $\Ph$ is linear we identify $T{\Ph}$ and $\Ph$.} $I:{\Ph}\to {\Ph}$ is a complex structure $I^2=-1$ which is compatible with the symplectic structure: $\omega(I(\X),I(\Y))= \omega(\X,\Y)$.
Such a complex structure determines a Hermitian form
\be
\bra \X,\Y\ket = H(\X,\Y) + i \omega(\X,\Y),
\ee
on $\Ph_{\mathbb{C}}$, 
and also an associated choice of 
complex coordinates. Our conventions are such that the Hermitian form is linear in its second argument, $\langle \X,I(\Y)\rangle=i\langle\X,\Y\rangle$. The complex coordinates are given by a map  from phase space to the complexified Lagrangian
$Z_\Omega: \Ph \to \tilde{L}_{\mathbb{C}}$,
with \[Z_\Omega:\X\mapsto z_\X = \Omega x - \tx,\] which we require to satisfy\footnote{If we write more generally $Z_\Omega(\X)_a=\tau_{aA}\X^A$, then (\ref{cplxcond}) implies $\tau I=i\tau$. Thus, in the chosen parameterization $\tau_{aA}=(\Omega_{ab},-\delta_a{}^b)$, if we write $\Omega$ in terms of real matrices as $\Omega=\Omega_1+i\Omega_2$ with $\Omega_2$ positive, the complex structure on $T\Ph$ has the form 
\[I=\begin{pmatrix}\Omega_2^{-1}\Omega_1&-\Omega_2^{-1}\cr \Omega_1\Omega_2^{-1}\Omega_1+\Omega_2&-\Omega_1\Omega_2^{-1}\end{pmatrix}.\]
} 
\beq\label{cplxcond}
Z_\Omega(I\X) =i Z_\Omega(\X).
\eeq
After a short calculation (writing $\Omega$ in terms of real matrices, $\Omega=\Omega_1+i\Omega_2$, $\Omega_2$ positive), the Hermitian form is found to be given by 
\beqn
 \bra \X,\Y\ket 
=  \bar{z}_\X{}^T\Omega_2^{-1} {z}_\Y.
\eeqn

Given a state $\Phi(x,\tx)$, it is convenient to recast it in the complex polarization as
\be
\Phi(x,\tx) = e^{i\pi\, x^T \Omega\, x}\, \Theta(z_\X,\bar{z}_\X). 
\ee
With this redefinition we can describe the scalar product 
as a Bargmann integral
\be
|\!|\Theta|\!|^2  =\int_{M_\Lambda} \rd^d z \rd^d \bar{z} \ e^{-2\pi\, x(z)^T \Omega_2x(z)}\ |\Theta(z,\bz) |^2.
\ee
where $x(z)= \Omega_2^{-1}{\rm Im}(z)$. From these expressions, it is clear that $Z_\Omega$ and its Hermitian conjugate act on $\Theta$ as creation and annihilation operators, satisfying the algebra
\be
\left[ \hat{Z}_{\Omega a},\hat{Z}^\dagger_{\Omega b} \right] = \frac{1}{\pi}\Omega_{2 ab}.
\ee
Starting from the Zak representation (\ref{Zakrep}) of $\hat\X$
on $\Phi$, one finds that they act on $\Theta$ as differential operators: 
\bea
 \hat{Z}_\Omega\to \left(\frac{1}{\pi} \Omega_2\pa_{\bar{z}}  \right)  ,\qquad 
\hat{Z}^\dagger_\Omega \to 
 -
 \left(\frac{1}{\pi}\Omega_2 \pa_z + z-\bar{z} \right).
\eea
%
%
From this expression it is clear that the vacuum state is a {\it holomorphic} functional which satisfies the quasi-periodicity condition (\ref{QP2})
\beq
\Theta(z+\Omega n-\tilde{n})
= 
e^{-i\pi\, n^T\Omega\, n}
e^{-2\pi i\, n^T z}\, \Theta(z)
\eeq

There is a {\it unique} holomorphic vacuum functional satisfying this condition given by the lattice theta function
\be
\Theta_\Lambda(z,\Omega)= \sum_{n\in \ell}  e^{i\pi\, n^T\Omega\, n} e^{2\pi i\, n^T z}.
\ee
We can now analyze what happens to the vacuum state when we modify the complex structure \cite{Axelrod:1989xt}, (holding $\omega$ and $\X$ fixed). 
Given the identification\footnote{To facilitate this calculation, note that this implies $\hat{E}_\Omega=\langle\X,\X\rangle=\bar{z}\Omega_2^{-1}z$ and under a variation in complex structure, we have $\delta z=\frac{1}{2i}\Omega_2\Delta (z-\bar{z})$.} $\hat{E}_\Omega=H(\X,\X)$, we write a variation of the complex structure in terms of $\Delta :=\Omega_2^{-1}\delta \Omega\ \Omega_2^{-1}$ 
and one finds that  
\be
\delta \hat{E}_\Omega = \frac1{2i}\left[\hat{Z}_a\bar{\Delta}^{ab}\hat{Z}_b - \hat{Z}^{\dagger}_a \Delta^{ab} \hat{Z}^\dagger_b \right].
\ee
This transformation can be reabsorbed into an infinitesimal Bogoliubov transformation. Two vacua are related to each other by squeezing 
$|0'\ket \simeq (1+\frac{i\pi}{4} Z^{\dagger} \Delta Z^\dagger+...) |0\ket\equiv  N \exp(i\pi Z^{\dagger} \Xi Z^\dagger) |0\ket$,
 which shows that the vacuum associated with the ``Hamiltonian" $\hat{E}_{\Omega'}$ contains an arbitrarily large number of $\hat{E}_{\Omega}$ excitations. 
This is reminiscent of the Unruh ambiguity inherent to the choice of vacuum for quantum fields in curved space, where different frames of reference (choices of time slicing) correspond to different vacua which are squeezed with respect to each other.
It is interesting that such an effect takes place in the context of modular  polarizations of space.

The geometrical reason behind this relies on the fact that the vacuum wave function
vanishes at a point $\mathbb{O}$, which can be thought as specifying a frame of reference for the quantum state. 
In the Schr\"{o}dinger representation this point represents $\infty$, the locus at which a wavefunction vanishes.  Since the modular cell is compact, the 
shape of the state around this point is accessible to observation, and the different vacua measure effectively in which quantum reference frame the system finds itself.
It is tempting to draw a speculative analogy with  the Unruh effect where the nature of the quantum vacuum depends on the accelerated observer's frame of reference. 
The analogy can be pushed further if one identifies the Unruh horizon with the point $\mathbb{O}$ which represents a point of unobservability, since probability vanishes at that point.

\section{Rotation and Translation Invariance in the Face of Discreteness}

\def\rotn{\mathbb{O}}

In this final section, we comment on how spatial symmetries are realized in modular quantizations. Ordinarily, one would say that a theory defined on a torus has only discrete versions of these symmetries. 
As we are about to see, this is not true. Even continuous rotations and translations are realized in a precise sense on a modular space. One can see that, given the Stone-von Neumann theorem, this is inevitable because any symmetry that exists in the Schr\"odinger representation ought to be realized in any other (unitarily equivalent) quantization. 

Before presenting our main line of argument, let us comment that resolving (a close relative of) this issue is of fundamental importance to quantum gravity. Indeed, the main conundrum for any theory of quantum gravity is to be able to find a way to reconcile 
the presence of a fundamental length and energy scale with the principle of relativity.  The basic issue is that the Lorentz contraction of length and the dilation of time for relative observers renders seemingly impossible the ability to have a scale on which all observers agree. 
This fundamental issue has attracted a lot of attention in recent years
and is still at the core of any attempt to produce a viable theory of quantum gravity.

A useful analogy in quantum mechanics is provided by angular momentum. Classically, the angular momentum describes a point on a sphere, and its canonical actions are given by rotations which are symmetries of the sphere. Quantum mechanically, representations are discrete; the eigenvalue of the angular momentum is evenly spaced and this can be interpreted as a discretization of the sphere, in which it is replaced by a  discrete set of circles equally spaced along the z-axis, say, a geometrical picture rigorously realized in the Kirillov orbit method \cite{Kirillov}. These orbits correspond to the weight lattice of the representation. The lattice apparently destroys the rotational symmetry of the classical description. But we know that is not true: quantum theory restores the rotational symmetry by arranging for a superposition of spin states which in turn parametrize a sphere's-worth of states. The spin states are merely a basis for the entire state space, and the basis is not invariant under rotations.
\myfig{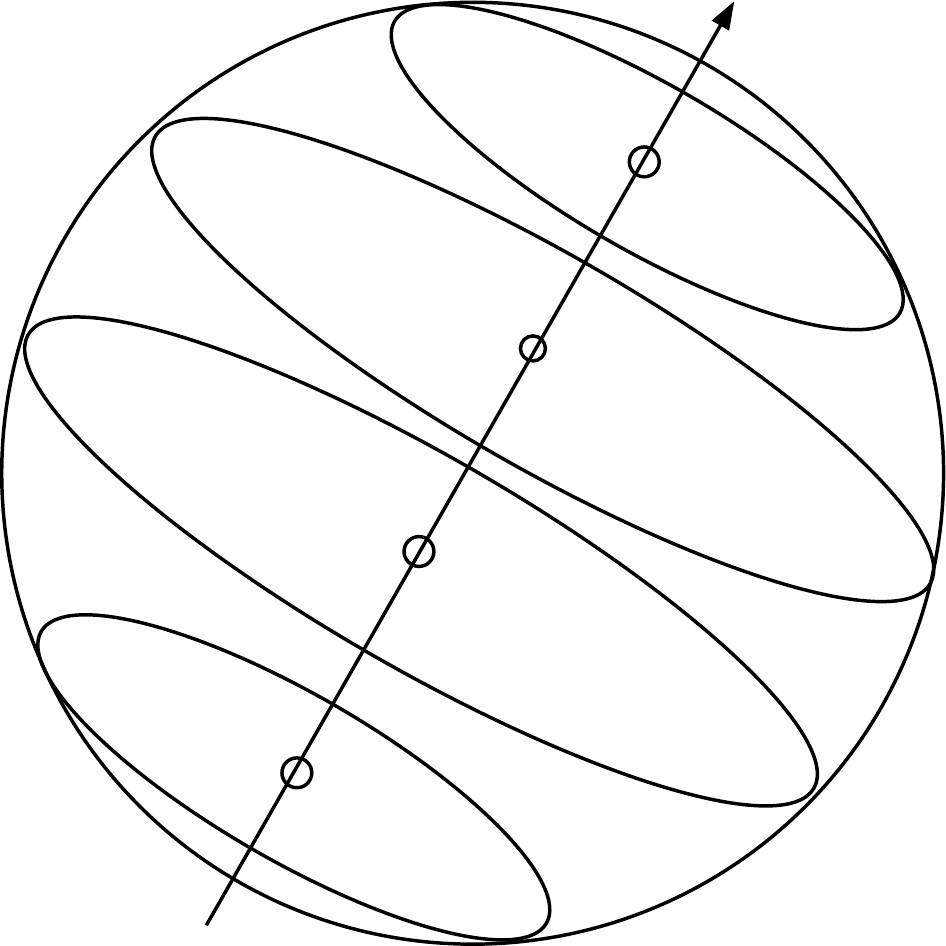}{4}{Discretization of sphere. }

So how does this analogy work in the context of modular space? In the Schr\"odinger representation, rotations act within the set of states $|x\rangle$, giving back a linear combination of these states. In the modular polarization, clearly, we should regard the rotations as acting on the basis, resulting in a new choice of modular cell, i.e., a new 'quantization axis'. Such a transformation corresponds to a canonical transformation.

Let us start again with the puzzle we are facing. As we have seen, once we introduce a fundamental scale, the natural polarization that respects the presence of this scale introduces a bilagrangian lattice $\Lambda=\ell\oplus \tilde{\ell}$ embedded in phase space $\cal P$, and  equipped with a neutral metric
$\eta$.  Moreover, we have seen that the Hilbert space is given by a space of sections of a line bundle $L_\Lambda$ over the torus $T_\Lambda = {\cal P}/\Lambda$ and that the embedding $T_\Lambda \hookrightarrow L_\Lambda$ is characterized by a lift, $\alpha_\eta \in {\rm U}(1)$. 
Rotations act on  phase space as symplectic transformations
\be
\X = (x,\tx) \mapsto \rotn\cdot\X = (O x, O^T \tx),\qquad O \in {\rm O}(d). 
\ee
These transformations not only preserve $\omega$, but they also preserve the polarization metric $\eta$ and the quantum metric $H$. Remarkably, we can show that rotations are in fact the {\it only} transformations that preserve the given Born geometry $(\omega, \eta,H)$ of the   modular space. In other words, we have that the rotation group lies at the intersection of the symplectic, neutral and doubly orthogonal groups,
\be\label{orthinter}
 {\rm O}(d)={\rm Sp}(2d) \cap {\rm O}(d,d) \cap {\rm O}(2d).
\ee
Here ${\rm Sp}(2d)$ is the symplectic group preserving $\omega$,
${\rm O}(d,d)$ is the neutral group preserving $\eta$ and ${\rm O}(2d)$ the orthogonal group preserving $H$. It is also interesting to understand the pairwise intersections. For instance, the group of transformations that preserves the symplectic structure and the polarization metric $\eta$ (hence the lift) is given by 
\be
  {\rm GL}(d)={\rm Sp}(2d) \cap {\rm O}(d,d).
\ee
Here GL$(d)$ is the general linear group that represents the sets of frames $e^a_i$ determining the vacuum metric $H(x,\tx)  = x^a q_{ab} x^b + \tx_a q^{ab} \tx_b $,
with $q^{ab} =  e^a_i \delta^{ij} e^a_j$ and $q_{ab} $  its inverse, while the space of vacuum metrics associated with a given Hilbert space is encoded in ${\rm GL}(d)/\rm{O}(d)$.
The other pairwise intersections are 
\be
{\rm U}(d)=  {\rm Sp}(2d) \cap {\rm O}(2d),\qquad {\rm O}(d)\times {\rm O}(d)
=  {\rm O}(d,d) \cap {\rm O}(2d).
\ee
The space ${\rm U}(d)/{\rm O}(d)$  represents the choice of inequivalent lifts $\eta$ that preserve the symplectic structure and the vacuum state, while the
quotient of the double orthogonal group by O$(d)$ represent the choice of inequivalent symplectic structures, hence Weyl groups,  that possess the same Hilbert space with the same vacuum.


The presence of a lattice structure naively breaks rotational invariance.
Indeed under a generic rotation, the lattice is mapped to a different lattice 
$ \rotn\Lambda \neq \Lambda$. It is only for a discrete subgroup which is a subset of  $O(d,\mathbb{Z}) $ for which $\rotn\Lambda =\Lambda$ that we expect the symmetry to be implemented exactly. 
For all other rotations, we expect the symmetry to be broken. The discrete lattice that follows from the introduction of a fundamental scale therefore  breaks the symmetry to a discrete subgroup.
This is exactly the conclusion one would reach if we were in the case in which  space-time is not modular, but this is not the right conclusion if  space is modular. The main simple but profound point is that a modular space does not come from the discretization of space but from the discretization of phase space and at the quantum level, phase space is non-commutative.
In order to show what this implies, recall that the lattice elements $\lambda =(n,\tn) \in \ell \oplus \tilde{\ell}$ are labels of  commutative observables 
\be
U_\lambda = e^{i\tilde{n}\cdot \hat{x}} e^{-i n\cdot\hat\tx} \in \hat\Lambda_{\eta},
\ee
where $\Lambda_\eta$ is a shortform for the abelian subgroup $\hat\Lambda_{\alpha_\eta}$ and $U$ is a morphism, $ U_{\lambda} U_\mu = U_{\lambda +\mu}$.
Now after a rotation, these operators are mapped onto another commutative subalgebra  $\hat{\Lambda}'_\eta $ with $\Lambda'=\rotn\Lambda$, still satisfying that $U_{\rotn\lambda} U_{\rotn\mu} = U_{\rotn(\lambda +\mu)}$ . 
The main point is however that these two commutative algebras {\it do not} commute with each other, $[U_\lambda , U_{\rotn\lambda}]\neq 0$. In fact we can express the non-commutation easily as 
\be 
U_\lambda U_{\rotn\lambda} = e^{2\pi i\omega(\lambda, \rotn \lambda)} U_{\rotn\lambda} U_{\lambda},
\ee
and unless the rotation is tuned to be such that all the angles $\omega(\lambda, \rotn\lambda)$ are integers, the non-commutativity is manifest.
Since these two sets of commuting observables do not commute with each other,  we can find a unitary transformation ${\cal U}_\rotn$ that maps one set onto the other.
\be
{\cal U}_\rotn U_\lambda {\cal U}_\rotn^\dagger = U_{\rotn\lambda}, \quad {\cal U}_\rotn^\dagger = {\cal U}_\rotn^{-1}, \quad \forall \lambda \in \Lambda.
\ee
When the rotation $\rotn$ can be written as an exponential of an infinitesimal transformation $O^a{}_b=(\exp \theta )^a{}_b$  where $\delta_{bc}\theta^{c}{}_a  =-\delta_{ac}\theta^{c}{}_b$, one may check that the unitary generator can be simply written in terms of the basic operators $\hat{x},\hat{\tx}$ as 
\be 
{\cal U}_\rotn=\exp (2\pi i\ \theta^a{}_b \hat{x}^b \hat{\tx}_a).
\ee
So indeed the rotation maps one modular space $T_{\Lambda}$ into another $T_{\rotn\Lambda}$, but since by construction, a modular space is a choice of polarization, any other modular space is related to the original one by a change of polarization, i.e., a canonical transformation.

This should be contrasted with the usual picture of discretizing space where 
the lattice and its image are associated with different spaces but where the algebra of functions on each space still commute with each other.
Because of this commutativity, it is not possible to relate them in any way and the two discretizations are not equivalent. 
This is also the case in the double field theory picture in string theory where the target space is doubled to include winding modes but the space-time coordinate $x$ and its dual, the winding coordinates $\tx$, are assumed to be commuting with each other (for a review, consult \cite{Zwiebach:2011rg}). In this picture where one regards the target as a doubled torus arising from compactification, the two sets of plane waves are totally unrelated and rotational symmetry is broken. This is one of the essential differences between the double space picture and the modular space picture in which different polarizations do not commute with each other.  
The main mechanism behind the symmetry restoration is the fact that 
a rotation does not act on the discrete space itself. It maps one discrete space to another isomorphic one. Since modular space is the quantum analog of a choice of a Lagrangian in phase space, the analogy is that after a rotation we generically end up in another quantum Lagrangian.
The property that there is a way to think about space as a Lagrangian in phase space and the fact that transformations change the Lagrangian is a characterization of {\it relative locality} \cite{AmelinoCamelia:2011bm}. Space is a notion relative to the observer, and different observers rotated with respect to one another experience isomorphic but different notions of space\cite{AmelinoCamelia:2011bm}, \cite{Freidel:2011mt}. The argument given here is closely related to a proposal for a resolution of the paradox in quantum gravity \cite{Rovelli:2002vp}, where it was argued that geometric operators associated with different boosted observers might not commute. The difference here is that we have a precise realization of this idea, which shows that it has to be associated with a realization of space as modular space. We will discuss the application of these ideas in quantum gravity and string theory elsewhere.

Now return to the analogy of the spin system, which goes as follows:  consider the quantization of a spin $N$ system, $[S_i,S_j]= \epsilon_{ijk}S^k$. The choice of a modular space $T_\Lambda$ is analogous to a choice of a Cartan subgroup in the rotation group, $S_z$ say. The quantization implies that the spectrum of $S_z$ is discrete, with $2N+1$ integer-spaced values.
Now the geometrical picture of the discretization of the sphere, which is a rigorous description of the quantum geometry associated with the observer $S_z$, seems to break rotational invariance. The resolution is due to the superposition principle of quantum mechanics:  after a rotation $S_z \to S_x$ say, eigenstates $|n\ket_z$ of $S_z$ are also superpositions of  eigenstates of $S_x$. The superposition $|n\ket_z = \sum_m U_{nm} |m\ket_x$ is realized by the unitary transformation $U^\dagger S_zU =S_x$. In other words states which are localized with respect to one polarization are delocalized with respect to another.

This mechanism of mapping localized discrete states into delocalized states expressed as a superposition is the key mechanism.
In order to appreciate this delocalization principle in the context of modular space,  we will formalize what we mean by a localized state.  
The map  ${\cal H}_{\Lambda} \to {\cal H}_{\rotn\Lambda}$ can be 
expressed as a composition $Z_{\rotn(\Lambda)} Z^{-1}_\Lambda$ of Zak transform an its inverse. It can be written as the integral transformation
\be
{\cal U}_\rotn \Phi_\Lambda(\X)= \int_{M_\Lambda} \rd \Y\ G_\rotn(\X,\Y) \Phi_\Lambda(\Y),
\ee 
where the integral kernel is  explicitely  given by 
\be
G_\rotn(\X,\Y)  = \sum_{(n,m)\in \ell^2} e^{-2\pi i( O(n)\cdot \tx) } \delta^{(d)} ( x+O(n) , y +  m) e^{2\pi i(m\cdot   \tilde{y})}.
\ee
This is the integral kernel for the map ${\cal H}_{\Lambda} \to {\cal H}_{\rotn\Lambda}$ given by
\be
{\cal U}_\rotn \Phi_\Lambda(\X)= \int_{M_\Lambda} \frac{\rd \Y}{{(2\pi)^d}}\ G_\rotn(\X,\Y) \Phi_\Lambda(\Y),
\ee 
where the integral is over a  fundamental cell.
The kernel belongs to $L_{\rotn(\Lambda)} \times L_{\Lambda}^*$ and it satisfies the property that 
$G_\rotn(\X+\lambda,\Y+\mu) =  e^{i\pi(\eta-\omega)(\lambda,\X)} 
G_\rotn(\X,\Y) e^{i\pi(\eta-\omega)(\X,\mu)}$ for $(\lambda,\mu)\in \Lambda^2$.
This means that this map is well-defined, independent of the choice of fundamental cell and mapping sections of $L_{\Lambda}$ into sections of $L_{\rotn\Lambda}$. It can also be checked to be unitary.
It is even simpler to describe this unitary map if one works with the complex polarization's functions $\Theta(z,\bar{z})$ described in the previous section. If we work with the canonical complex structure $\Omega_{ab} = i\delta_{ab}$, we have that 
\be
 {\cal U}_\rotn \Theta_\Lambda(\X)= \Theta_\Lambda(O^T\dd z, O^T\dd \bar{z}).
\ee
This shows the simplicity of the complex polarizations; the unitary map is simply represented as a rotation of the holomorphic coordinates.

Now that we have understood how the rotational symmetry is restored, let's look at translation symmetry. Since the lattice possesses a preferred origin, naively one expects only a subset of discrete translations $T_{\mathbb{O}}$ to preserve the lattice. At first sight the situation seems radically different because under a translation $T_{\mathbb{O}} \X=\X +\mathbb{O}$, or in components $(x,\tx)\to (x+o,\tx +\tilde{o})$, the commutative algebra $\hat{\Lambda}$ is mapped onto itself and therefore the algebras  $\hat{\Lambda} $ and $T_{\mathbb{O}}\hat\Lambda$ commute with each other.
There are however two elements that come to our rescue and insure that translations, like rotations, are implemented canonically. The first one is the fact that the translation algebra itself is not commutative as we have seen in a previous section. The second is that even if the commutative algebra is the same under translation, the lift $\alpha_{(\eta,\mathbb{O})}$, and hence the line bundle $L_{(\eta,\mathbb{O})}$ that defines the Hilbert space, is not the same. We have seen that a lift is characterized in general by a choice of a polarization metric $\eta$ and by a characteristic $\mathbb{O}\in {\cal P}$. A translation will change the characteristic and gives $T_{\mathbb{O}}L_{(\eta,0)} =L_{(\eta,-\mathbb{O})}$.
A change of characteristic corresponds to a change of phase of the commutative operators, which can be reabsorbed by a unitary transformation
\be
 U_\lambda(\X +\mathbb{O})= T_{\mathbb{O}} U_\lambda = e^{2\pi i \omega(\lambda,\mathbb{O}) } U_\lambda = W_{\mathbb{O}}^\dagger U_\lambda W_{\mathbb{O}},
\ee
where $W_{\mathbb{O}} $ denotes the Weyl translation operator (\ref{Weyl}).
We see that once again a translation  that changes the lattice  can be implemented via a unitary isomorphism as
it maps a state $\Phi(\X)$ into  
\be
W_{\mathbb{O}} \Phi(\X) = e^{i\pi\omega(\X,\mathbb{O})} \Phi(\X+\mathbb{O}).
\ee
The fact that translations act by a phase shift is the underlying reason
for  localization.

One can ask indeed why it is that we do not experience a mixture of space and momentum space as modular space is suggesting?
In principle it should be possible to prepare a state which is a superposition of $\Phi$ and its translation, that is 
\be
\Psi(\X) = \Phi(\X) + W_{\mathbb{O}} \Phi(\X)= \Phi(\X) +e^{i\pi\omega(\X,\mathbb{O})} \Phi(\X+\mathbb{O}).
\ee
If one could keep the coherence of that state we could in principle observe  interferences between a state and its translation along $\mathbb{O}$
due to the presence of the phase $\omega(\X,\mathbb{O})$.
Now if $\X$ and $\mathbb{O}$ belong to the same Lagrangian, this phase vanishes and no interference can be experienced.
Note that one has to take $\mathbb{O} $ inside a Lagrangian  complementary to $\X$ in order to witness interference.
Decoherence on the other hand, due to the imprecision of the state preparation and the interaction of the states with an external system that kicks the state around, will ensure that states which are translated along orthogonal Lagrangians decohere with each other. The reduced density matrix will not mix states that are translated into transverse Lagrangians.
This can explain why we seem to live in a classical Lagrangian instead of a modular one.

\subsection{Extensification: Many is Large!}
In this final section, we investigate in what sense we can recover in some approximation the usual notion of space starting from the quantum modular perspective.  
So far we have discussed the Hilbert space for one unit of flux and we have seen that it is associated with a Planckian cell of area $\hbar$.  We also have seen in the previous section how geometrical transformations that change the shape of the lattice can be implemented as unitary transformations.
The question we want  to investigate now  is  what happens if we ``extensify'' the modular cell, which is of  unit Planckian area, into a cell of area $N $ in Planck units?  More precisely, we consider {\it changing} the lattice in a variety of ways which changes the size of its fundamental cell. This is a non-trivial process, as it changes the structure of the theory, and in particular modifies the Hilbert space.
More precisely, while a change of the lattice shape $\Lambda \to \Lambda'$ that doesn't modify  its size can be implemented unitarily,
a change that changes its size  leads to inequivalent Hilbert spaces ${\cal H}_\Lambda $ and ${\cal H}_{\Lambda'}$. The fact that they are inequivalent can be seen in the fact that they generically have different degeneracies in the vacuum sector. Fundamentally this is due to the fact that modular spaces depend on a  fundamental scale. This should be contrasted with the Schr\"odinger representation for which a rescaling of space $\varphi(q) \to \varphi(Nq)$ is a diagonal unitary transformation.

Given a bilagrangian lattice 
$\Lambda =\ell \oplus \tilde{\ell}$ there are really two different ways in which we can extend the cell. We can extensify  modular space  along the Lagrangian $L$ if we replace the lattice by $N\ell \oplus \tilde{\ell}$. Or we can extensify modular space  along $\tilde{L}$ by  performing the rescaling $\Lambda \to  \ell \oplus \tilde{N}\tilde{\ell}$.
 It will be convenient to introduce the short-hand notations 
 \be
 N\dd\Lambda\dd \tilde{N} :=  (N\ell) \oplus (\tilde{N}\tilde{\ell}), \qquad N\dd\Lambda:=N\dd\Lambda\dd 1,\qquad \Lambda \dd \tilde{N} = 1\dd\Lambda \dd\tilde{N}.
 \ee 
 With this notation we distinguish $N\dd\Lambda$ which corresponds to 
 an extensification along $L$ from $(N\Lambda)= N\dd \Lambda \dd N$
 which corresponds to a ``coarsening" operation\footnote{We use the term `coarsening' to refer to a uniform rescaling of the size of the phase space lattice. This should be distinguished from a coarse-graining operation, which would preserve the volume of the cell.}.
 The first operation makes the size of space relatively bigger than the size of momentum, while the second increases the size of the cell homogeneously in both directions.  
 It will be important to note that there are two equivalent ways to formalize the process of extensification: we can either  expand the lattice $\Lambda \to N\dd\Lambda$, while keeping the fundamental scales $(\lambda,\varepsilon)$ fixed  or we can  equivalently keep the lattice fixed 
 while  rescaling  the fundamental scales as 
 $
 \lambda \to   \lambda/N$, and  $ \varepsilon \to \varepsilon$.
 This corresponds to a rescaling $(G,\hbar)\sim (\lambda/\varepsilon, \lambda\varepsilon/2\pi )\to (G/N,\hbar/N)$.  The large $N$ limit corresponds to  a limit in which both $\hbar$ and the gravitational coupling (used here as a conversion factor) are sent to zero at the same speed. It is reminiscent of the relative locality limit considered in \cite{AmelinoCamelia:2011bm}.
\myfig{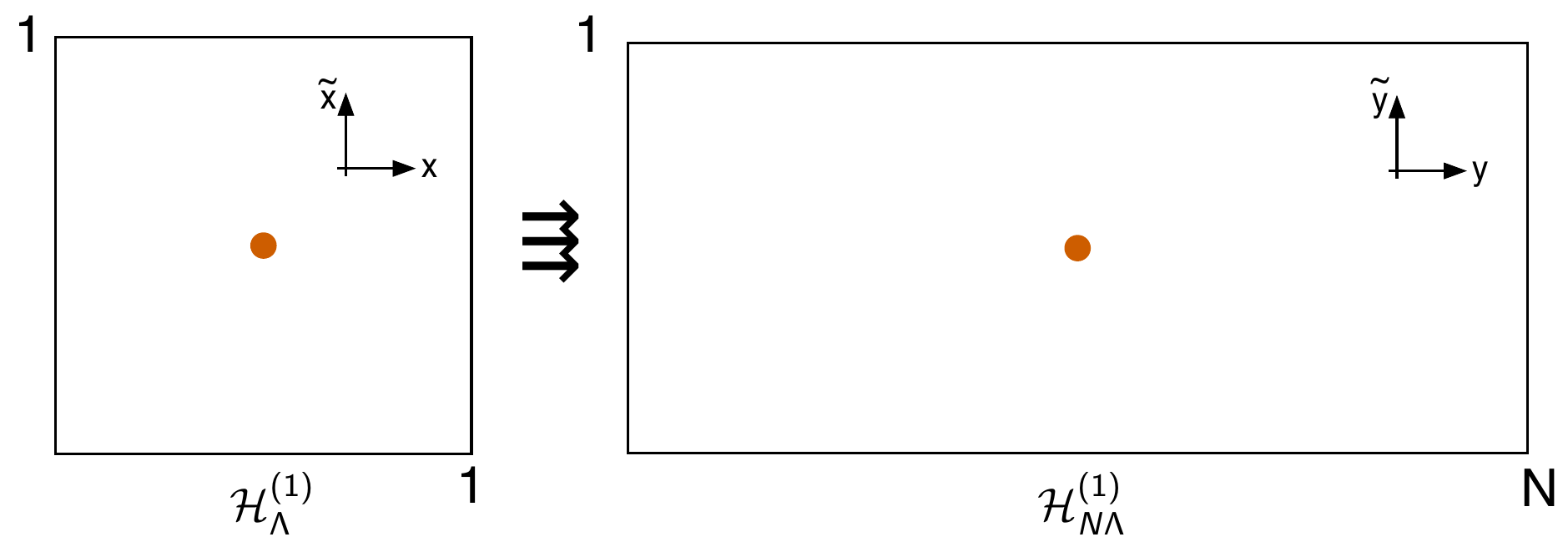}{12}{Rescaling the lattice by $N$ results in a {\it new} Hilbert space.}

The coarsening limit on the other hand corresponds to a rescaling 
 $\lambda \to   \lambda/N$, $\varepsilon \to \varepsilon/N$ of both fundamental scales. This corresponds to the transformation
$(G,\hbar)\to (G,\hbar/N^2)$, which apparently can be interpreted as a classical GR limit (once we introduce a curved version of modular space).
 From this perspective both extensification and coarsening  correspond to expanding towards a larger  phase space cell. 
While both contain a classical rescaling of $\hbar$,
 what distinguishes them is that during extensification we change the shape of the  cell, which is elongated along the space direction,  while during coarsening, we do not. 
 This is the first  main point: extensification corresponds  to an enlargement of the phase space cell, together with a proportional  increase of the fundamental tension. 

We also want to understand   what happens to the concept of   modular  space  when we consider multi-flux states ${\cal H}_{\Lambda}^{\otimes N}$,  instead of ${\cal H}_{\Lambda}$.
An interesting fact is that this is fundamentally related to the notion of extensification presented above. The Hilbert space of multi-flux  states
 can be decomposed into the product of the Hilbert space for the `center of mass' times the Hilbert space of  the `relative motion' around the center of mass
 \be
{\cal H}_\Lambda^{\otimes N} = {\cal H}_{\Lambda, \rm{com}}^{(N) }\otimes {\cal H}_{\Lambda,\rm{rel}}^{(N)}.
 \ee
 We can then show that the Hilbert space for the center of mass ${\cal H}_{\Lambda, \rm{com}}^{(N) }$ 
 is isomorphic to the extensified modular space ${\cal H}_{ {N}\Lambda}
$.   
Moreover the states obtained by extensification ${\cal H}_{ {N}\cdot \Lambda} \hookrightarrow
 {\cal H}_\Lambda^{\otimes N}$  can be embedded into the multi-flux Hilbert space as a subset of states which are highly coherent  with respect to the relative degrees of freedom. 
 This relation is in fact very natural:
since each flux in ${\cal H}_{\Lambda}$ occupies an elementary cell, a collection of $N$ fluxes occupies in a precise  sense a larger phase space cell.
What differentiates an extensified state possessing an emergent geometrical  interpretation, from another quantum state, is the way the $N$ fluxes are arranged within the cell. 
In order for the usual notion of space to appear, the elementary fluxes have to be arranged to lie regularly along a classical Lagrangian.

\newcommand{\extmap}{{\rm\widehat{Ex}}}

Let us first study the extensification process $\Lambda \to N\dd \Lambda$.
 The first thing  to establish is an understanding of  the difference between the Hilbert space ${\cal H}_{\Lambda}={\cal H}_{\Lambda}^{(1)}$ and its sections, versus those of the extensified version 
${\cal H}_{N\cdot\Lambda}^{(1)}$. 
 The superscript notation, which we have not used previously, now emphasizes that the sections are meant to be homogeneous of degree one and here we generalize this notion. 
As we have seen, ${\cal H}_{\Lambda}^{(1)}$ is\footnote{ Strictly speaking the Hilbert space and the line bundle  depend on the choice of lift and we should refer to it as ${\cal H}^{(1)}_{(\Lambda,\eta)} $ and $L_{(\Lambda,\eta)}$ respectively. Since we keep $\eta$ fixed in this section we drop the extra label for notational clarity.} the space $\Gamma^{(1)}(L_{\Lambda})$ of degree one sections of the line bundle $L_{\Lambda} \to T_\Lambda$, where we recall that the line bundle is defined by  $(u,\X+\lambda ) \sim(ue^{i\pi (\eta-\omega)(\lambda,\X)},\X) =(ue^{2i\pi n\tx},\X)$ and the degree $N$ of a section is the homogeneity degree $\Phi(\lambda u,\X)= \lambda^N \Phi(u,\X)$.
We can similarly define the Hilbert space ${\cal H}_{\Lambda}^{(N)} :=
\Gamma^{(N)}(L_{\Lambda})$ of degree $N$ sections. The degree measures the amount of symplectic flux going through a fundamental cell. If the phase space is of dimension $2d$ we have that 
\be
\int _{\mathbb{M}_\Lambda} \omega^d =N^d.
\ee

We now want to express that at the quantum level there are two equivalent ways of describing the extensified Hilbert space. We can either define it to be 
the Hilbert space  ${\cal H}_{N\cdot\Lambda}^{(1)}$ associated with a bigger lattice  or we can define it as the Hilbert space ${\cal H}_{\Lambda}^{(N)}$, associated with degree $N$ sections on the initial lattice. This second description corresponds to a rescaling of the fundamental constants.
Before describing the isomorphism, let us specify each space:
if we denote the element of the lattice by $( n,\tilde{n}) \in \ell \oplus \tilde{\ell}$, the elements of the extensified lattice are $(Nn,\tn) \in N\dd\Lambda$.
The elements $\Psi$  of ${\cal H}_{N\Lambda}^{(1)}$ satisfy
\beq
\Psi(y+Nn,\ty+\tilde{n})=e^{2\pi iNn\ty}\,\Psi(y,\ty),
\eeq
where we have introduced the natural coordinates for the extensified cell, $y\in [0,N), \ty\in [0,1)$.
On the other hand, the elements $\Phi$  of ${\cal H}_{\Lambda}^{(N)}$ 
are sections of $L_\Lambda$ satisfying
\be
\Phi(x+n,\tx+\tilde{n})=e^{2\pi N in\tx}\,\Phi(x,\tx),
\ee
where $(x,\tx)\in [0,1)\times [0,1)$ describe the reference cell.
The  isomorphism between these two descriptions is simply given by  the extensification mapping 
\bea
\extmap_N: {\cal H}^{(1)}_{N\cdot\Lambda}      &\to&  {\cal H}_{\Lambda}^{(N)}, \\
\Phi( y,\tilde{y}) &\to& (\extmap_N\Phi)(x,\tilde{x}):= \Phi\left( N{x}, \tilde{x}\right),
\eea
\noindent and indeed given $\Phi\in {\cal H}_\Lambda^{(1)}$, we have 
\beqn
(\extmap_N\Phi)(x+n,\tilde{x}+\tilde{n})&=&
\Phi\left( N(x+n), \tilde{x}+\tilde{n}\right)
=e^{2\pi iNn\tx}(\extmap_N\Phi)(x,\tilde{x}).
\eeqn

\myfig{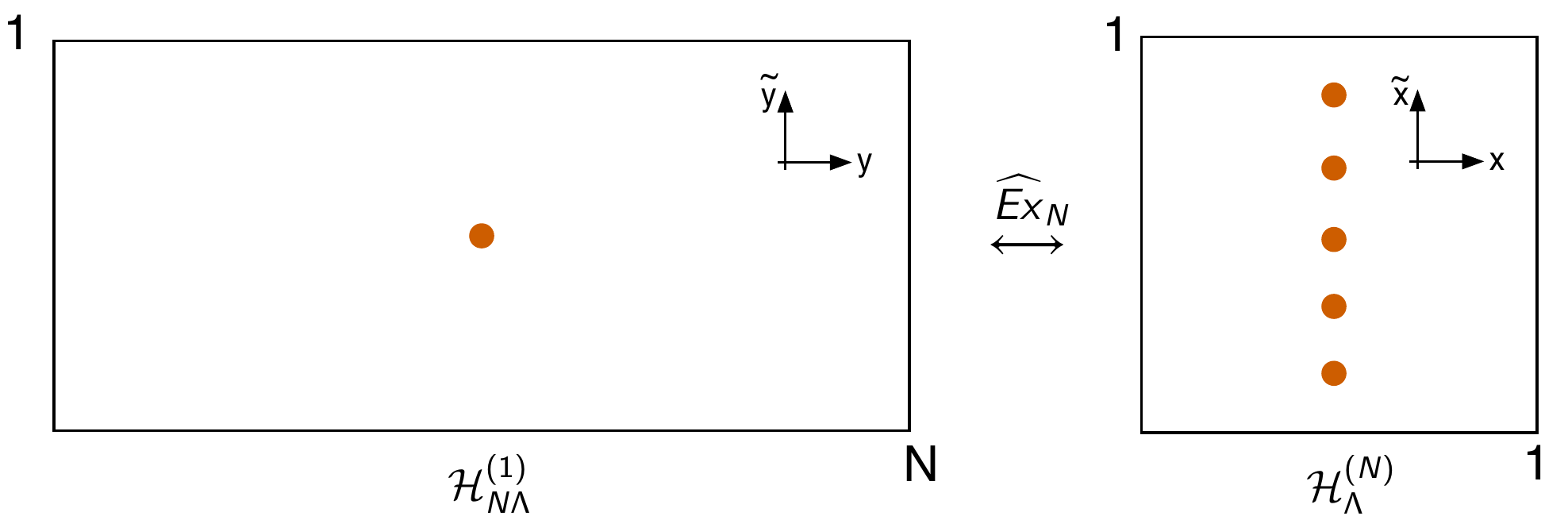}{12}{Pictorial view of the $\extmap_N$ isomorphism of Hilbert spaces. Degree $N$ sections have $N$ zeroes within a fundamental cell.}

In order to appreciate what is changing under extensification, one looks at the vacuum sector. As we have seen in  section (\ref{modv}), the choice of a ``Hamiltonian" $\hat{E}_\Omega= \hat{z}^\dagger\Omega_2^{-1} \hat{z}$ can be understood in terms of a choice of complex structure, which for simplicity here, we restrict to the simplest choice $\Omega_{ab}=i  \delta_{ab}$ for the lattice $\Lambda$. 
That is we assume 
\be
z= ix-\tx = i\tfrac{y}{N} -\tilde{y}. 
\ee
In order to analyze the vacuum sector and its excitations
we consider the complex polarization, 
and we define 
\be 
(\extmap_N\Phi)(x,\tx) =  e^{ - N \pi (x\cdot x)} \Theta(z,\bz),
\ee
where the contraction is now with the Kronecker matrix $x\cdot x =x^a\delta_{ab}x^b$ and the factor $N$ in the exponent takes care of the fact that $\Phi$ is a degree $N$ section.
As we have seen, the vacuum states are the holomorphic sections 
while the degree $N$ condition means that they transform  as \be\label{quasip}
 \Theta^{(N)}_{\Lambda}(z+ in +  \tn)
 = e^{ \pi N  (n \cdot n)} e^{- 2i \pi N  (n\cdot z)} \Theta^{(N)}_{\Lambda}(z).
\ee
The general solution of this equation is easy to find: one first expands $\Theta(z)$ in Fourier modes $ e^{2i\pi n\cdot z}$ and translates the previous equation into a recursion equation for the Fourier coefficients. One finds that there are $N^d$ independent  solutions labeled by 
$k\in (N \ell)/\ell $ and given by  
\be
\Theta^{(N)}_{(\Lambda,k)}(z)= \sum_{n \in \ell }e^{-\pi N \left( n +{k}/{N}\right)^2 } e^{2i\pi N ( n +{k}/{N})\cdot z }.
\ee
The number of independent functions that correspond to vacuum states is given by the total flux  $\int \omega^d$ over a fundamental cell of the extensified lattice. They are a generalization of Landau levels to $d$ dimensions \cite{Laughlin}. 
The important point is that they form a basis of ${\cal H}^{(N)}_\Lambda$ which is of dimension $N^d$. 

As we have seen the key property of the vacuum functional is the location of its zeroes. The  flux determines the number of zeroes: we expect to find $N$ zeroes of order $d$ per cell. In order to find them we write 
\be
\Theta^{(N)}_{(\Lambda,k)}(z)= e^{\pi N { x^2} } \sum_{n \in \ell }e^{-{\pi}{ N} \left( n -{k}/N+ {x} \right )^2 } e^{-2i\pi N ( n -{k}/{N})\cdot \tilde{x}  }.
\ee
For simplicity, we assume from now on that $\ell=\mathbb{Z}^d$ and therefore $k=(k^1,\cdots k^d)\in N\ell/ \ell $ can be represented by choosing  $k^a \in \{0,\cdots N-1\}$.
From this expression we see that the points  
\be
x=\left(\tfrac{1}2, \cdots,\tfrac12\right)  +\frac{k}{N},\qquad \tilde{x} =\left(\tfrac{1}2, \cdots,\tfrac12\right) +  \frac{ m}{N},\qquad m \in N\tilde{\ell}/ \tilde{\ell},
\ee
are zeroes of order $d$ of $\Theta^{(N)}_{(\Lambda,k)}$ labeled by elements of the quotient lattice  $N\tilde{\ell}/ \tilde{\ell}$. 
This follows from the fact that for these values the exchange $n^a \to -n^a-1$ changes the summand by a sign.
We see  that a general extensified state is a state with a particular coherency in the structure of its zeroes. We also see that the zeroes of the extensified vacuum states $\Theta_{(\Lambda,k)}^{(N)}$ are aligned along a Lagrangian manifold given by the condition $x:=(z-\bar{z})/2i=k/N$ ( see Fig. \ref{fig:extensify.pdf}). The choice of Lagrangian is controlled by the choice of complex structure that enters the definition of the ``Hamiltonian".
The coherency of these states is manifest in the fact that the zeroes are equally spaced  with regular spacing $\Delta \X \in N^{-1} \tilde{\ell}$ along the dual  Lagrangian $N^{-1} \tilde{\ell} = (N\ell)^\perp$.

An important fact is that the position of the zeroes  determines the vacuum wave-function, up to a constant. This follows from the fact that the ratio of two wave-functions with the same zeroes is analytic and periodic and hence constant. Using this fact, we can 
introduce another basis of ${\cal H}_{\Lambda}^{(N)}$ whose zeroes are distributed arbitrarily. One first considers the  theta function $\Theta_\Lambda(z)=\Theta ^{(1)}_{\Lambda}(z)$ which is the unique vacuum state of ${\cal H}_{\Lambda}^{(1)}$ and we define 
 \be
\Theta_{\Lambda}^{(N)}(z; z_i) :=  e^{-2i\pi N{c}\cdot (z- z_c) } \prod_{i=1}^N \Theta_{\Lambda} (z - z_i),
 \ee
 where we have introduced the `center of mass' position: $  z_c := (\sum_{i=1}^N z_i)/ N = ic -\tilde{c}$. The normalization is chosen so that it is translation covariant
 \be\label{transa}
  \Theta_{\Lambda}^{(N)}(z+a; z_i +a ) = e^{-2i\pi N{\rm Im}(a)(z-z_c)}\,\Theta_{\Lambda}^{(N)}(z; z_i).
 \ee
 \myfig{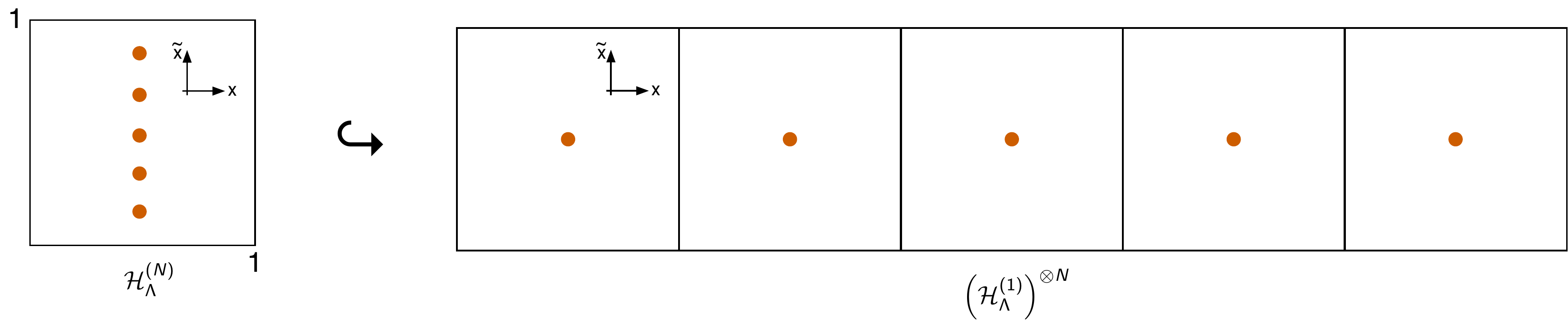}{14}{Organizing the zeroes of the vacuum wavefunction in ${\cal H}_\Lambda^{(N)}$ in a coherent fashion identifies an embedding ${\cal H}_\Lambda^{(N)}\hookrightarrow \left({\cal H}_\Lambda^{(1)}\right)^{\otimes N}$. }

This means that a change of the center of mass can be reabsorbed, up to an exponential factor, by a translation in $z$.
When  $a \in \Lambda$ we know that $ \Theta_{\Lambda}^{(N)}$ is mapped onto itself.  This means that  the  translation of the center of mass position  by an element of $ \Lambda$ is trivial.

This wave-function possesses zeroes at the points $z= \tfrac{i}2+\tfrac12 + z_i\, $, for $i=1,\cdots, N$.
The quasi-periodicity condition of this function reads 
\be
\Theta_{\Lambda}^{(N)} (z+in +\tn; z_i) = e^{2i\pi N(n\tilde{c}-\tn c)} 
e^{\pi N n^2} e^{-2i\pi n  N \cdot z} \Theta_{\Lambda}^{(N)} (z; z_i) .
\ee
This shows that it is in ${\cal H}_{\Lambda}^{(N)}$ provided the center of mass 
$(c,\tilde{c})$ belongs to the lattice $( N^{-1}\Lambda)$. When this is the case  the phase factor is equal to the identity for all $(n,\tn)\in \Lambda$.
Using the translation covariance we can  therefore label this function in terms of N elements $z_i$ such that $\sum_i z_i =0$  and an element 
$(c,\tilde{c})\in \Lambda/(N^{-1}\Lambda)$.

Let us finally comment on the physical relevance of the coherency of the position of zeroes of the wave function. 
First let us remark that the vacuum states 
$|\Theta_{z_i}\ket$  represented by the holomorphic functional $\Theta_\Lambda(z-z_i)$ are coherent states centered around 
$\bra \hat{\X}_i \ket=({\rm Im}(z_i), {\rm Re}(z_i))$, where  we  have introduced  the expectation value of $\X_i$ in the state $\Theta_{z_i}$ by $\bra\hat{\X}_i \ket$. The  dispersion of the states defined by $(\Delta \X)^2= \bra \Theta| (\hat{\X}- \bra\hat\X\ket)^2 |\Theta\ket $ is given by $\Delta \X=( 1/\pi,1) $ in the units we have chosen. Now a basic fact about the action of Hermitian operators on states is that it can be decomposed as \cite{aharonov2008quantum, Anandan:1990fq}
\be
\hat{\X}_i^A |\Theta_{z_i}\ket = \bra\hat{\X}_i \ket  |\Theta_{z_i}\ket  +\Delta \X^A_i
 |\Theta_{z_i}\ket^\perp ,
\ee
where $ |\Theta_{z_i}\ket^\perp$ is a normalized state orthogonal to $ |\Theta_{z_i}\ket$. What is of interest to us is how this decomposition scales when we consider a multi-particle state $|\Theta^{(N)}\ket := |\Theta_{z_1}\ket \otimes \cdots \otimes |\Theta_{z_N}\ket$. We define the center of mass observables $\hat{\X}=\tfrac1{N} \sum_i \hat{\X}_i$ and a corresponding orthogonal normalized state 
$|\Theta^{(N)}\ket^\perp = \tfrac1{\sqrt{N}} \sum_i |\Theta_{z_1}\ket \otimes \cdots \otimes |\Theta_{z_i}\ket^\perp   \cdots \otimes |\Theta_{z_N}\ket $. 
Now, since all individual states have the same coherency $\Delta \X_i= \Delta \X$,  we can evaluate
\be
\hat{\X}  |\Theta^{(N)}\ket=  \bra \hat{\X} \ket |\Theta^{(N)}\ket + \frac{ \Delta \X }{\sqrt{N}}  |\Theta^{(N)}\ket^\perp.
\ee
From here we can see that there exist two radically different physical situations. If we assume that the expectation values $\bra \hat{\X}_i \ket$ are randomly distributed, along a random walk, with diffusion constant $D$  say, then we know that in the large $N$ limit we have that $ \bra \hat{\X} \ket \sim D /\sqrt{N}$, since we are summing over random variables. Therefore, for an incoherent state the average value of an observable is of the same order as  its dispersion in the large $N$ limit and it cannot be interpreted classically.
If, on the other hand, the distribution of the average value is coherently distributed 
we can see that the average value $\bra \hat{\X}\ket$ is of order one, while 
the dispersion is of order $1/\sqrt{N}$, and the state becomes classical in the large $N$ limit.
In order to get the ``classicalization'' of a state we have used, we need the quantum coherency of individual states and the classical coherency of the distribution of average values. Under these conditions we  conclude that, quite paradoxically, coherency is necessary in order to obtain the usual classical limit.

As a conclusion, let us remark that the relationship between an extensified wave-function and the superposition of multi-flux wave-functions extends to the coarsening operation. If we assume that $\Lambda =\mathbb{ Z}^d\oplus \mathbb{Z}^d$ we can investigate the basic coarsening operation that amounts to the rescaling $\Lambda \to 2\Lambda$.
This operation can be realized as an embedding ${\cal H}_{2\cdot \Lambda\cdot 2} \hookrightarrow {\cal H}_{\Lambda}^{\otimes 4}$.
This embedding is encoded by  the Riemann identity \cite{Mumford1}:
\be
\Theta_\Lambda\left(2{z}\right) =\frac1{2^d \Theta_\Lambda(0)^3}
\sum_{(a,b)\in (\mathbb{ Z}/ 2\mathbb{ Z})^{2d}}
e^{4\pi a^2} e^{2i\pi a z} [\Theta_\Lambda(z +ia+b)]^4.
\ee
In this section we have shown that modular space has properties which are far 
more intricate than those of the usual classical space. Now we have a possibility to realize  states associated with larger space as multi-flux states.
These identifications in principle blur the line between what is pure geometry and what can be considered as matter.
This unification of matter and geometry contained in modular space is one of the most fascinating aspects
of this discussion.

\bigskip

\section{Discussion}

In this paper, we have introduced the notion of quantum polarization, corresponding to a generic choice of a maximally commutative sub-group of the Heisenberg group, found in ordinary non-relativistic quantum mechanics. This construction can be thought of as the proper mathematical underpinning of modular variables \cite{Aharonov}, which have been used to best describe purely quantum phenomena (such as double slit experiments and their brethren). Within such an application, the lattice (and consequently a length scale) associated with the maximally commutative sub-group is set contextually by some experimental apparatus (such as a slit spacing). 

More generally, we interpret the construction to give rise to a purely quantum notion of space. In this interpretation, we take the scale associated with the quantum polarization to be fundamental. The usual classical notion of space is recovered from this quantum space in a singular limit\footnote{
A typical example of a singular limit is the limit of zero viscosity of a viscous fluid.  A singular limit is a mathematically consistent description, but it contains an impossibility realized in the non-singular limit: as is well known in an Euler fluid, planes can't fly.  Other examples of singular limits are the geometric optics limit of wave optics, when the
wavelength is neglected, the thermodynamic limit of
statistical mechanics, when the number of particles is assumed to be infinite, and finally, the classical physics limit of
quantum physics.}  \cite{Berry} in which the 
 fundamental length scale decouples. This interpretation is certainly more conjectural, but it seems natural to suppose that the fundamental length scale is associated with the gravitational (or string) scale, $ \varepsilon/\lambda \sim c^4/G_N$. In this context, we might think of the unit flux modular space as a `space-time bit'. 
The principle of equivalence together with the relativity principle postulate that the gravitational tension is independent of the nature of constituents. 
It is interesting to notice that in physical units the gravitational tension is huge, of order $10^{17} kg/\AA$. Consequently, if we do make this identification, in most experiments we can safely ignore the presence of a fundamental scale, as fluctuations along momentum directions are negligible compared to fluctuations in position directions. In making this identification, we are suggesting that the quantumness of space is in fact quantum gravitational. Hence one might interpret this as the {\it gravitization of the quantum}. 
 Some of our reasons for believing this conjecture are contained in earlier publications, and forthcoming papers will provide further and more focussed evidence.  

A commutative sub-group of the Heisenberg group is determined by an 
integral self-dual lattice. We have shown that geometrical structures arise, principally a symmetric bilinear form $\eta$ of split signature
and a positive definite form $H$. Remarkably, these are precisely the structures that arise in generalized geometry \cite{Gualtieri:2003dx} as well as in the T-duality symmetric formulation of string theory that we refer to as metastring theory. 
It is quite possible that the structure of modular quantizations has observable consequences in metastring theory, through UV-IR mixing, when the appropriate limit is taken to obtain local effective field theories in large space-times.
%
 
Another contribution presented in this paper is a resolution of a central conundrum in all formulations of quantum gravity concerning how to make the existence of a fundamental invariant gravitational scale consistent with the relativity principle. There are two issues here: the first is that discretizations break continuous symmetries, while length scales are typically not invariant under Lorentz boosts. Even in ordinary quantum mechanics we are concerned with the former. Indeed the modular polarization, with its associated lattice, apparently breaks the continuous rotational and translational symmetries of phase space. We have shown that this is a naive conclusion however, precisely because phase space is a non-commutative space, following a mechanism first proposed in \cite{Rovelli:2002vp}. A choice of quantum polarization is a choice of basis, unitarily equivalent to any other. To reconcile these ideas, we must retool our thinking in terms of the {\it superpositions} of quantum spaces. This should be thought of as an implementation of the idea of relative locality \cite{AmelinoCamelia:2011bm}. We anticipate that the use of these ideas in quantum gravity will have similar consequences.

Our discussion suggests that the modular basis is rather special in ordinary quantum mechanics, because it seems to capture its essential non-locality. We also have seen that in the process of extensification the boundary between purely geometrical degrees of freedom and matter-like degrees of freedom encoded in the fluxes  gets blurred. This may be one of the most fascinating aspects of modular space and this is one of the questions that we hope to explore in a future work.
We will also address the restoration of Lorentz symmetry and the question of how causality is reconciled with non-locality. 
We also expect that 
the modular basis should find wider uses in foundational questions of quantum theory, including the problem of measurement, and the emergence of classical physics as a singular limit  of the underlying quantum world.
Finally it should allow us to bring a new perspective on the geometry of string theory since it is in this context that it first appeared \cite{Freidel:2015pka}.

\medskip\noindent {\bf Acknowledgements:}
Research supported in part by the U.S. Department of Energy under contracts DE-SC0015655 (RGL) and DE-FG02-13ER41917 (DM) and by the Government of Canada through NSERC and by the Province of Ontario through MRI (LF).
We thank Banff International Research Station for generous hospitality and for providing an inspiring environment.

\providecommand{\href}[2]{#2}\begingroup\raggedright\endgroup
\end{document}